\documentclass[11pt]{article}

\usepackage{amsmath,amssymb,ulem,graphicx}
\usepackage[margin=1in]{geometry}
\usepackage[sort&compress,numbers]{natbib}

\usepackage{hyperref}

\hypersetup{
colorlinks=true,urlcolor=blue,
	linkcolor = blue,
        urlcolor  = blue,
        citecolor = blue,
        anchorcolor = blue,
}
\hypersetup{pdftitle={Optical Potential on Lattice},pdfsubject={Manuscript}} 

\begin{document}
\def\beq{\begin{equation}}
\def\eeq{\end{equation}}
\def\bea{\begin{eqnarray}}
\def\eea{\end{eqnarray}}
\def\eq{\begin{eqnarray}}
\def\en{\end{eqnarray}}

\begin{titlepage}

\vspace{1cm}
\begin{center}{\Large\bf The Optical Potential on the Lattice}

\vspace{0.5cm}
\today

\vspace{0.5cm}
Dimitri Agadjanov$^{a}$, 
Michael D\"oring$^{b,c}$, 
Maxim Mai$^a$, 
Ulf-G. Mei{\ss}ner$^{a,d}$
and Akaki Rusetsky$^a$

\vspace{2em}
\footnotesize{\begin{tabular}{c}
$^a\,$ Helmholtz-Institut f\"ur Strahlen- und Kernphysik (Theorie) and\\
Bethe Center for Theoretical Physics,\\
$\hspace{2mm}$   Universit\"at Bonn, D-53115 Bonn, Germany\\[2mm]
$^b$
Institute for Nuclear Studies and APSIS, Department of Physics, The George Washington University,\\ 725 21$^{\rm st}$ St. NW, Washington, DC 20052, USA\\[2mm]
$^c$
Thomas Jefferson National Accelerator Facility, 12000 Jefferson Ave, Newport News, VA 23606, USA\\[2mm]
$^d\,$
Institute for Advanced Simulation (IAS-4), Institut f\"ur Kernphysik 
(IKP-3) and\\ J\"ulich Center for Hadron Physics,
Forschungszentrum J\"ulich, D-52425 J\"ulich, Germany
\end{tabular}  }

\vspace{1cm}

\begin{abstract}
\noindent
The extraction of hadron-hadron scattering parameters from lattice data by using
the L\"uscher approach becomes increasingly complicated in the presence of
inelastic channels. We propose a method for the direct extraction of the
complex hadron-hadron optical potential on the lattice, which does not require
the use of the multi-channel L\"uscher formalism. Moreover, this method is
applicable without modifications if some inelastic channels contain three or
more particles. 
\end{abstract}

\vspace{1cm}
\footnotesize{\begin{tabular}{ll}
{\bf{Pacs:}}$\!\!\!\!$& 03.65.Nk, 11.80.Gw, 12.38.Gc
\\
{\bf{Keywords:}}$\!\!\!\!$& Lattice QCD, inelastic channels,
non-relativistic EFT, L\"uscher equation
\\
\end{tabular}}
\end{center}
\end{titlepage}
\setcounter{page}{2}

\section{Introduction}\label{sec:intro}

The L\"uscher approach~\cite{Luescher-torus} has become a standard tool to study
hadron-hadron scattering processes on the lattice. The use of  this approach in
case of elastic scattering is conceptually straightforward: besides technical
complications, caused by partial-wave mixing, each measured energy level at a
given volume uniquely determines the value of the elastic phase shift at the
same energy.

In the presence of multiple channels, the extraction of the scattering phase 
becomes more involved. In case when only two-particle coupled channels appear,
one can make use of the coupled-channel L\"uscher
equation~\cite{Lage-KN,Lage-scalar,He,Sharpe,Briceno-multi,Liu,PengGuo-multi} 
and fit a simple pole parameterization for the  multi-channel $K$-matrix
elements to the measured energy spectrum in the finite
volume~\cite{Wilson-pieta}. A more sophisticated parameterization of the
$K$-matrix elements, which is applicable in a wider range of the energies, can
be obtained using unitarized chiral perturbation theory
(ChPT)~\cite{oset,Doring-scalar,oset-in-a-finite-volume}. This approach has been
successfully applied, e.g., in Ref.~\cite{Wilson-rho} to analyze 
coupled-channel $\pi\pi-K\bar K$ $P$-wave scattering and to study the properties
of the $\rho$-meson. Note the  difference to the one-channel case: here, one has
to determine several  $K$-matrix elements (unknowns) from a single measurement
of a finite-volume energy level. Hence, using some kind of
(phenomenology-inspired)  parameterizations of the multi-channel $K$-matrix
elements becomes inevitable in practical applications.

In case when some of the inelastic channels contain three or more particles, the
situation is far more complicated. Despite the recent progress in the
formulation of the theoretical
framework~\cite{Polejaeva,Briceno-3,Sharpe-3,Rios}, it is still too cumbersome
to be directly used in the analysis of the data.  Moreover, the problem of the
choice of the parameterization for three-particle scattering might become more
difficult (and lead to even  larger theoretical uncertainties) than in
two-particle scattering.

From the above discussion it is clear that a straightforward extension of the 
L\"uscher approach through the inclusion of more channels has its  limits that
are reached rather quickly. On the other hand, many interesting systems, which
are already studied on the lattice, may decay into multiple channels. In our
opinion, the present situation warrants a rethinking of the paradigm. One may
for example  explore the possibility to analyze the lattice data without
explicitly resolving the scattering into each coupled channel separately. Such a
detailed information is usually  not needed in practice. Instead, in the
continuum scattering problem, the effect of inelastic channels could  be
included in the so-called optical potential~\cite{Feshbach,Kerman:1959fr}, whose
imaginary part is non-zero due to the presence of the open inelastic channels.
In many cases, it would be sufficient to learn how one extracts the real and
imaginary parts of the optical potential from the lattice data, without
resorting to the multi-channel L\"uscher approach. In the present paper, we
propose such a method, which  heavily relies on the use of twisted
boundary
conditions~\cite{Bedaque,Sachrajda,rest-twisted,Chen,Agadjanov-twisted}. Due to
this, the method has its own limitations, but there exist certain systems,
where  it could in principle be applied. In particular, we have  the following
systems in mind:

\begin{itemize}
\item
The scattering in the coupled-channel $\pi\eta-K\bar K$
system in the vicinity of the $K\bar K$ threshold and the $a_0(980)$ resonance.
\item
The spectrum and decays of the $XYZ$ states; namely, 
$Z_c(3900)^\pm$ that couples to the channels $J/\psi\pi^\pm$, $h_c\pi^\pm$ and
$(D\bar D^*)^\pm$ (this system was recently studied in Ref.~\cite{Padmanath})
or the $Z_c(4025)$ that couples to the $D^*\bar D^*$ and $h_c\pi$ channels
(see, e.g., Ref.~\cite{ChuanLiu}).
\end{itemize}

There certainly exist other systems where this method  can be used. It should
also be stressed that the systems, where  the partial twisting
(i.e., twisting only the valence quarks) can be carried out, are interesting in
the first place -- for an obvious reason.  All examples listed above belong to
this class.  In general, the partial twisting can always be carried out when the
annihilation diagrams are absent. In the presence of annihilation diagrams, each
particular case should be analyzed separately, invoking the methods of effective
field theories in a finite  volume~\cite{Agadjanov-twisted}. The present paper
contains an example of such an analysis.

Further, note that there exists an alternative method for the extraction  of
 hadron-hadron interaction potentials from the measured
Bethe-Salpeter  wave functions on the Euclidean lattice. This method  goes under
the name of the HAL QCD approach and its essentials are explained in
Refs.~\cite{HAL-essentials,HAL-derivatives}.  Most interesting in the present
context is the  claim that the HAL QCD approach can be extended to the
multi-channel systems, including the channels that contain three and
more-particles~\cite{HAL-multi}. It should also be pointed out
that this approach has already been used to study various systems on the
lattice, including the analysis of coupled-channel baryon-baryon
scattering (see, e.g., Ref.~\cite{HAL-applications}). It would be interesting to
compare our method with the HAL QCD approach.

The layout of the present paper is as follows. In Sect.~\ref{sec:complex_plane}
we discuss the theoretical framework for the extraction of the real and
imaginary parts of the optical potential and provide an illustration of the
method with synthetic data, generated by using unitarized ChPT. Further, in
Sect.~\ref{sec:realisticpseudophase}, we discuss the role of  twisted boundary
conditions for measuring the optical potential. Namely, the possibility of
imposing partially twisted boundary conditions is explored  in
Sect.~\ref{sec:partialtwisting}. Here, we also discuss the possibility of
imposing the different boundary conditions on the quark and antiquark fields. 
The analysis of  synthetic data, including an error analysis,  is presented in
Sect.~\ref{sec:simulation}. Finally, Sect.~\ref{sec:concl} contains our
conclusions.


\section{Optical potential in the L\"uscher approach}
\label{sec:complex_plane}

\subsection{Multichannel potential, projection operators}

In the continuum scattering theory, the inelastic channels can be effectively 
included in the so-called optical potential by using the Feshbach projection
operator technique~\cite{Feshbach}. Namely, let us start from the multi-channel
$T$-matrix which obeys the Lippmann-Schwinger equation 
\begin{align}\label{eq:LS} T=V+VG_0T\,. \end{align} Here, $V$ is the potential
and $G_0=(E-H_0)^{-1}$ denotes the free  Green's function with $E$ the total
energy in the center-of-mass system. The quantities $T,V,G_0$ are all $N\times
N$ matrices in channel space.

In case when only two-particle intermediate states are present, using
dimensional regularization together with the threshold expansion,  it can be
shown that the Lippmann-Schwinger equation~\eqref{eq:LS} after 
partial-wave  expansion reduces to an algebraic matrix equation (see, e.g.,
Ref.~\cite{Lage-dist}). With the proper choice of normalization, the matrix
$G_0(E)$ in this case takes the form
\begin{align}
G_0(E)&=\mbox{diag}\,(ip_1(E),\cdots,ip_n(E))\,,
\end{align}
where $p_k(E)$ denotes the magnitude of the center-of-mass three-momentum, i.e.,
\begin{align}\label{eq:pcms}
p_k(E)&=\frac{1}
{2E}\sqrt{\left(E^2-\left(m_1^{(k)}+m_2^{(k)}\right)^2\right)
\left(E^2-\left(m_1^{(k)}-m_2^{(k)}\right)^2\right)}
\end{align}
and $m_{1,2}^{(k)}$ are the masses of particles in the $k^{\rm th}$ scattering 
channel. Hence, if dimensional regularization is used in case of   two-particle
channels, the potential $V$ coincides with the multi-channel $K$ matrix. The
latter quantity can always be defined,
irrespectively of the used regularization. Our
final results are of course independent of the use of a particular
regularization.

Suppose further that we focus on the scattering in a given two-particle channel.
Let us introduce the projection operators $P$ and $Q=1-P$, which project on this
channel and on the rest, respectively. In
the following, we refer to them as the primary (index $P$) and the secondary
(index $Q$) channels.  The secondary channels may contain an arbitrary number of
particles.  It is then straightforward to show that the quantity $T_P(E)=PT(E)P$
obeys the  following single-channel Lippmann-Schwinger equation
\begin{align}
T_P(E)&=W(E)+W(E)G_P(E)T_P(E)\,,
\end{align}
where 
\begin{align}
&W(E)=P\biggl(V+VQ\frac{1}{E-H_0-QVQ}QV\biggr)P \quad
\text{ and }\quad G_P(E)=PG_0(E)P\,.
\end{align}

It is easily seen that, while $V$ is Hermitean, $W(E)$ above the secondary 
threshold(s) is not. The imaginary part of $W(E)$ is expressed through the 
transition amplitudes into the secondary channels
\begin{align}
W(E)-W^\dagger(E)&=-2\pi i\, PT_Q^\dagger(E)Q\,\delta(E-H_0)\,QT_Q(E)P\,,
\end{align}
where 
\begin{align}
T_Q(E)&=V+VG_Q(E)T_Q(E) \quad\text{ and }\quad G_Q(E)=QG_0(E)Q\,.
\end{align}

For illustration, let us consider  scattering in the $\pi \eta - K\bar K$
coupled channels. Let $K\bar K$ and $\pi\eta$ be the primary and secondary
channels, respectively. Then, the formulae for the S-wave scattering  take the
following form (we suppress the partial-wave indices for brevity):
\begin{align}
T_{K\bar K\to K\bar K}(E)=W(E)+ip_{K\bar K}\,W(E)T_{K\bar K\to K\bar K}(E)\,.
\end{align}
Here,
\begin{align}
W(E)=V_{K\bar K\to K\bar K}+\frac{ip_{\pi\eta}V_{K\bar K\to \pi\eta}^2}{1-ip_{\pi\eta}V_{\pi\eta\to \pi\eta}}\, ,
\end{align}
$p_{K\bar K}$, $p_{\pi\eta}$ denote the magnitude of the relative three-momenta 
in the center-of-mass  frame in the respective channel, as given in
Eq.~\eqref{eq:pcms}.

It is often useful to introduce the so-called $M$-matrix $M=V^{-1}$. In terms of
this quantity, the above formula can be rewritten in the  following
form:
\eq
W^{-1}(E)=M_{K\bar K\to K\bar K}
-\frac{M_{K\bar K\to \pi\eta}^2}{M_{\pi\eta\to\pi\eta}-ip_{\pi\eta}}\, .
\label{winv}
\en
Using the latter form can be justified, when a resonance
near the  elastic threshold exists that shows up as a pole on the real
axis in $V$.  In contrast, the quantity $M$ is smooth in
this case and can be  Taylor-expanded near threshold.

In a finite volume, one may define a counterpart of the scattering amplitude
$T_{K\bar K\to K\bar K}(E)$. Imposing, e.g., periodic boundary conditions leads
to the modification of the loop functions
(for simplicity, we restrict ourselves to the S-waves from here on and
neglect partial wave mixing)
\eq
ip_k\to \frac{2}{\sqrt{\pi} L}\,Z_{00}(1;q_k^2) \quad\text{for}\quad
q_k=\frac{p_kL}{2\pi}\, ,
\en
whereas the potential $V$ remains unchanged up to exponentially suppressed 
corrections. In the above expressions, $L$ is the size of the cubic box and
$Z_{00}$ denotes the L\"uscher zeta-function.

The energy levels of a system in a finite volume coincide with the poles of the
modified scattering amplitude. The position of these poles is determined from
the secular equation
\eq\label{eq:secular}
\biggl(M_{K\bar K\to K\bar K}-\frac{2}{\sqrt{\pi}L}\,
Z_{00}(1;q_{K\bar K}^2)\biggr)
\biggl(M_{\pi\eta\to\pi\eta}-\frac{2}{\sqrt{\pi}L}\,Z_{00}(1;q_{\pi\eta}^2)
\biggr)
-M_{K\bar K\to\pi\eta}^2=0\, .
\en
The positions of these poles on the real axis are the quantities that are
measured on the lattice.


\subsection{Continuation to the complex energy plane}

\begin{figure}[t]
\begin{center}
\includegraphics[width=\linewidth]{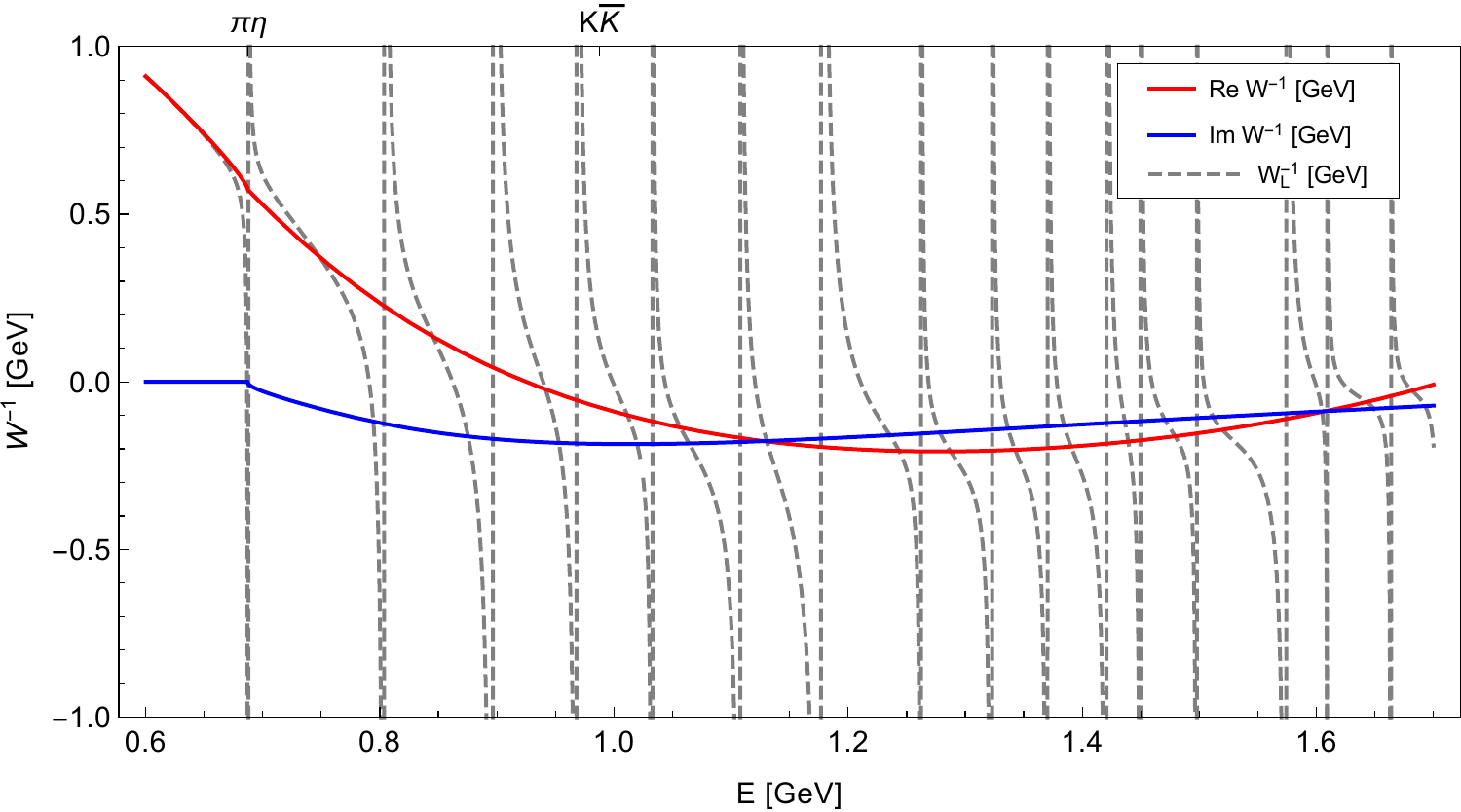}
\caption{The real and imaginary parts of the quantity $W^{-1}(E)$, as well as
its finite-volume counterpart $W_L^{-1}(E)$ for $L=5M_\pi^{-1}$.}
\label{fig:pseudophase-exact}\end{center}
\end{figure}

\begin{figure}[t]
\begin{center}
\includegraphics[width=\linewidth]{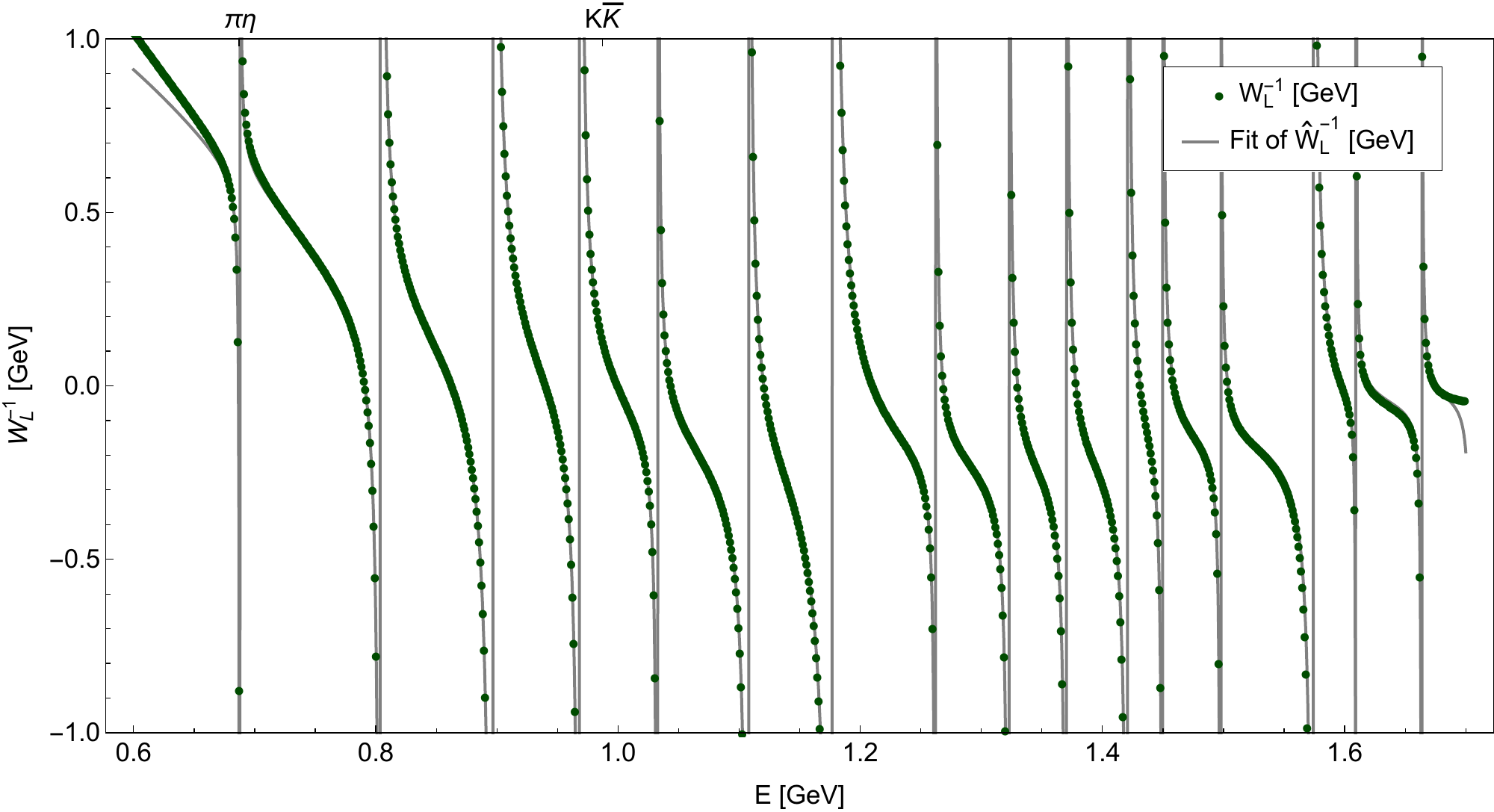}
\end{center}
\caption{Fit of the function specified in Eq.~\eqref{eq:realaxis} to the 
quantity $W_L^{-1}(E)$ for $L=5M_\pi^{-1}$ and uniformly distributed values of energy $E$.}
\label{fig:pseudophase-exact-fit}
\end{figure}

\begin{figure}[t]
\begin{center}
\includegraphics[width=0.98\linewidth]{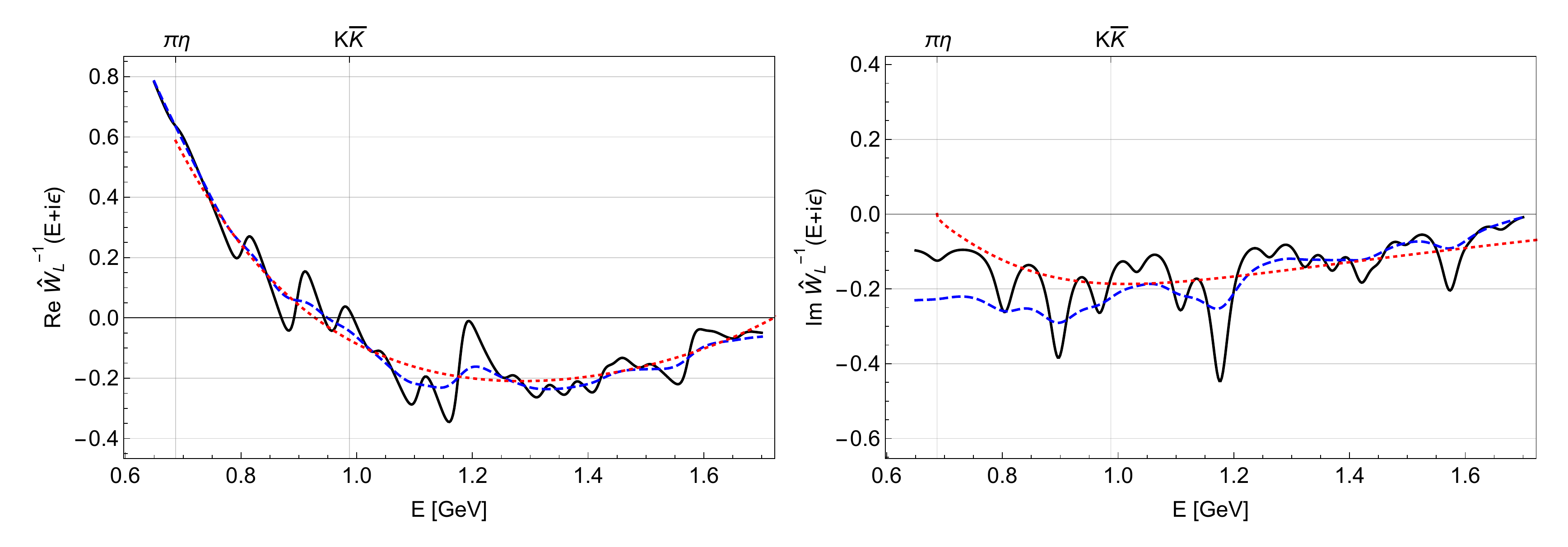}
\end{center}
\caption{The real and imaginary parts of the quantity $\hat W_L^{-1}(E+i\varepsilon)$
for $\varepsilon=0.02~\mbox{GeV}$ (solid black lines) and 
$\varepsilon=0.05~\mbox{GeV}$ (dashed blue lines)
versus the real and imaginary parts of the infinite-volume counterpart 
$W^{-1}(E)$ (dotted red lines). All quantities are given in units of GeV.}
\label{fig:oscillations}
\end{figure}

The main question, which we are trying to answer, can now be formulated  as
follows: Is it possible to extract the real and imaginary parts of $W(E)$ from
the measurements performed on the lattice? We expect that the answer  exists and
is positive, for the following reason. Let us imagine that all scattering
experiments in Nature are  performed in a very large hall with certain boundary
conditions imposed  on its walls. It is {\it a priori} clear that
nothing could change in the interpretation of the results of this experiment, if
the walls are moved to  infinity. Consequently, there {\it should} exist a
consistent definition of the infinite-volume limit in a finite-volume theory
that yields all quantities defined within the scattering theory  in the
continuum. Since the optical potential is one of these, there should exist a
quantity defined in a finite volume, which coincides with the optical potential
in the infinite-volume limit.
 
In order to find out, which quantity corresponds to the optical potential in a
finite volume and how the infinite-volume limit should be performed,  let us
follow the same pattern as in the infinite volume. Namely, we apply the
one-channel L\"uscher equation for the analysis of data, instead of the
two-channel one. As a result, we get:
\eq\label{eq:WL}
W_L^{-1}(E):=
\frac{2}{\sqrt{\pi}L}\,Z_{00}(1;q_{K\bar K}^2)=
M_{K\bar K\to K\bar K}
-\frac{M_{K\bar K\to \pi\eta}^2}{M_{\pi\eta\to\pi\eta}
-\frac{2}{\sqrt{\pi}L}\,Z_{00}(1;q_{\pi\eta}^2)}\, .
\en
The left-hand side of this equation is measured on the lattice at fixed values
of $p_{K\bar K}$, corresponding to the discrete energy levels in a  finite
volume. Methods to measure $W_L^{-1}$ are discussed in
Sect.~\ref{sec:realisticpseudophase}.  The quantity on the right-hand side is
proportional to the  cotangent of the so-called {\it pseudophase}, defined as
the phase extracted with  the one-channel L\"uscher
equation~\cite{Lage-KN,Lage-scalar}. It coincides with the usual scattering
phase in the absence of secondary channels. 

Fig.~\ref{fig:pseudophase-exact} shows the real and imaginary parts of the
quantity $W^{-1}(E)$ that is  constructed by using a simple parameterization  of
the two-channel $T$-matrix, based on unitarized ChPT (see  Ref.~\cite{Oller}).
For comparison, the finite-volume counterpart  $W_L^{-1}(E)$, which is defined
by Eq.~(\ref{eq:WL}), is also shown. If the secondary channels were absent,
$W^{-1}(E)$ would be real and equal to $W_L^{-1}(E)$, up to exponentially
suppressed contributions. Fig.~\ref{fig:pseudophase-exact} clearly demonstrates
the effect of neglecting the secondary channels. While the ``true'' function
$W^{-1}(E)$ is a smooth (and complex) function of energy, the (real)  function
$W_L^{-1}(E)$ has a tower of poles and zeros. The (simple)  zeros of
$W_L^{-1}(E)$ (poles of $W_L(E)$) emerge, when $E$  coincides with one of the
energy levels in the interacting $\pi\eta$ system.  The background, obtained by
subtracting all simple poles, is a smooth function of $E$. It should be 
stressed that this statement stays valid even in the presence of  multiple
secondary channels, some of which containing three or more particles. The only
singularities that emerge in general are the  simple poles that can be traced
back to the eigenvalues of the total  Hamiltonian restricted to the subspace of
the secondary states\footnote{Strictly  speaking, this argument applies only to
$W_L(E)$. However, assuming the  absence of accidental multiple zeros in
$W_L(E)$, one may extend this argument to $W_L^{-1}(E)$.}.

It is important to note that, if $L$ tends to infinity, the optical potential
does not have a well-defined limit at a given energy. As the energy levels in
the secondary channel(s) condense towards the threshold, the quantity
$W_L^{-1}(E)$ at a fixed $E$ oscillates from $-\infty$ to $+\infty$. Thus, the
question arises, how the quantity $W^{-1}(E)$ can be obtained  in the
infinite-volume limit.

It should be pointed out that this question has been already addressed in the
literature in the past. In this respect, we find Ref.~\cite{DeWitt} most useful.
In this paper it is pointed out that, in order to give a correct causal
description of the scattering process, one should consider adiabatic switching
of the interaction. This is equivalent to  attaching an infinitesimal imaginary
part $E\to E+i\varepsilon$ to the energy. Further, as argued in
Ref.~\cite{DeWitt}, the limits $L\to\infty$ and $\varepsilon\to 0$ are not
interchangeable. A correct infinite-volume limit is obtained, when  $L\to\infty$
is performed first (see Ref.~\cite{finite-t} for a more detailed discussion of
this issue). Physically, this statement is clear. The quantity  $\varepsilon$
defines the available energy resolution, and the distance between the
neighboring energy levels tends to zero for $L\to\infty$. If this distance
becomes smaller than the energy resolution, the discrete levels merge into a cut
and the infinite-volume limit is achieved. It is also clear, why the
infinite-volume limit does not exist on the real axis: $\varepsilon=0$ 
corresponds to an infinitely sharp resolution and the cut is never observed.

The above qualitative discussion can be related to L\"uscher's regular summation
theorem~\cite{Luescher-regular}. On the real axis above threshold, the 
zeta-function $Z_{00}(1;q_{\pi\eta}^2)$ in Eq.~(\ref{eq:WL}) does not have a
well-defined limit. Assume, however, that the energy $E$ gets a small positive
imaginary part, $E\to E+i\varepsilon$. The variable $q_{\pi\eta}^2$ also becomes
imaginary:
\eq
q_{\pi\eta}^2\to
q_{\pi\eta}^2+\frac{i\varepsilon E}{2}\,\biggl(\frac{L}{2\pi}\biggr)^2
\biggl(1-\frac{(M_\eta^2-M_\pi^2)^2}{E^4}\biggr)=
q_{\pi\eta}^2+i\varepsilon'\, .
\en
It is immediately seen that above threshold, $E>M_\eta+M_\pi$, the quantity
$\varepsilon'$ is strictly positive. Now, for real energies $E$,  the nearest
singularity  is located  at the distance $\varepsilon$ from the real axis, so
the regular summation theorem can be applied. It can be straightforwardly
verified that the remainder term in this theorem vanishes as
$\exp(-\varepsilon'L)$ (modulo powers of $L$), when $L\to\infty$.

The above argumentation can be readily extended to the cases when intermediate
states contain any number of particles. Consider a generic loop diagram in the
effective field theory where these particles appear as internal lines. It is
most convenient to use old-fashioned time-ordered perturbation theory, where
the integrand contains the energy denominator ${(E+i\varepsilon-w_1({\bf p}_1)
\ldots -w_n({\bf p}_n))^{-1}}$. Here, $w_i({\bf p}_1)\, ,~i=1,\ldots,n$ stand
for the (real) energies of the individual particles in the intermediate state.
It is clear that, if $\varepsilon\neq 0$, the denominator never  vanishes, and
the regular summation theorem can be applied. The remainder,  as in the
two-particle case, vanishes exponentially when $\varepsilon\neq 0$.

The analytic continuation into the complex plane can be done as follows. 
Suppose one can measure the quantity $W_L^{-1}(E)$ on the real axis. Bearing in 
mind the above discussion, one may fit this function by a sum of simple poles
plus a regular background.
Fig.~\ref{fig:pseudophase-exact-fit} shows the  result of such a fit which was
performed by using the function \eq\label{eq:realaxis} \hat
W_L^{-1}(E)=\sum_i\frac{Z_i}{E-Y_i}+D_0+D_1E+D_2E^2+D_3E^3\, \en to fit a sample
of the exact $W_L^{-1}$ without errors. The exact values of the fit parameters
are not listed here  since Fig.~\ref{fig:pseudophase-exact-fit} is given for
the illustrative purposes only. In the actual numerical  simulation of
Sect.~\ref{sec:simulation},  the order of the polynomial is varied.

The continuation into the complex plane is trivial: one uses
Eq.~(\ref{eq:realaxis}) with fixed values of $Z_i,~Y_i,~D_i$  and makes the
substitution $E\to E+i\varepsilon$.  The real and imaginary parts of the
quantity $\hat W_L^{-1}(E+i\varepsilon)$ for  $\varepsilon=0.02~\mbox{GeV}$ and
$\varepsilon=0.05~\mbox{GeV}$ are shown in  Fig.~\ref{fig:oscillations}. For
comparison, the real and imaginary parts of the  infinite-volume counterpart
$W^{-1}(E)$ are also given. As seen, the finite-volume  ``optical potential''
oscillates around the true one and the magnitude of such oscillation  grows
larger, when $\varepsilon$ becomes smaller. On the other hand, the artifacts 
caused by a finite $\varepsilon$ grow, when $\varepsilon$ becomes large.


\subsection{Infinite-Volume Extrapolation}\label{sec:smearing}

From the above discussion it is clear that,  performing the limit $L\to\infty$
for a fixed  $\varepsilon$, and then taking $\varepsilon\to 0$, the
infinite-volume limit is restored  from $\hat W_L^{-1}(E+i\varepsilon)$. For the
actual extraction on the lattice, however,  taking the large volume limit could
be barely feasible.  An alternative to this procedure is to ``smooth'' the
oscillations arising  from Eq.~(\ref{eq:realaxis}) if evaluated at complex
energies at a finite $L$ and $\varepsilon$. This allows one to  perform the
extraction of the optical potential at a reasonable accuracy even at 
sufficiently small values of $L$. As in the present study the true optical
potential  is known, the validity of this procedure can be tested. We would like
to stress that  $LM_\pi=5$ used in this study is rather small and thus not
completely beyond reach.

In the present section we test two different algorithms for smoothing the
quantity  $\hat W_L^{-1}(E+i\varepsilon)$. In both cases, the result is called
$\hat W^{-1}$, i.e., the estimate  of the true infinite-volume potential
$W^{-1}$. The final results of the numerical  studies, presented in
Sect.~\ref{sec:simulation} are evaluated with both methods.

\subsubsection*{Parametric method}

\begin{figure}[t]
\begin{center}
\includegraphics[width=0.48\textwidth]{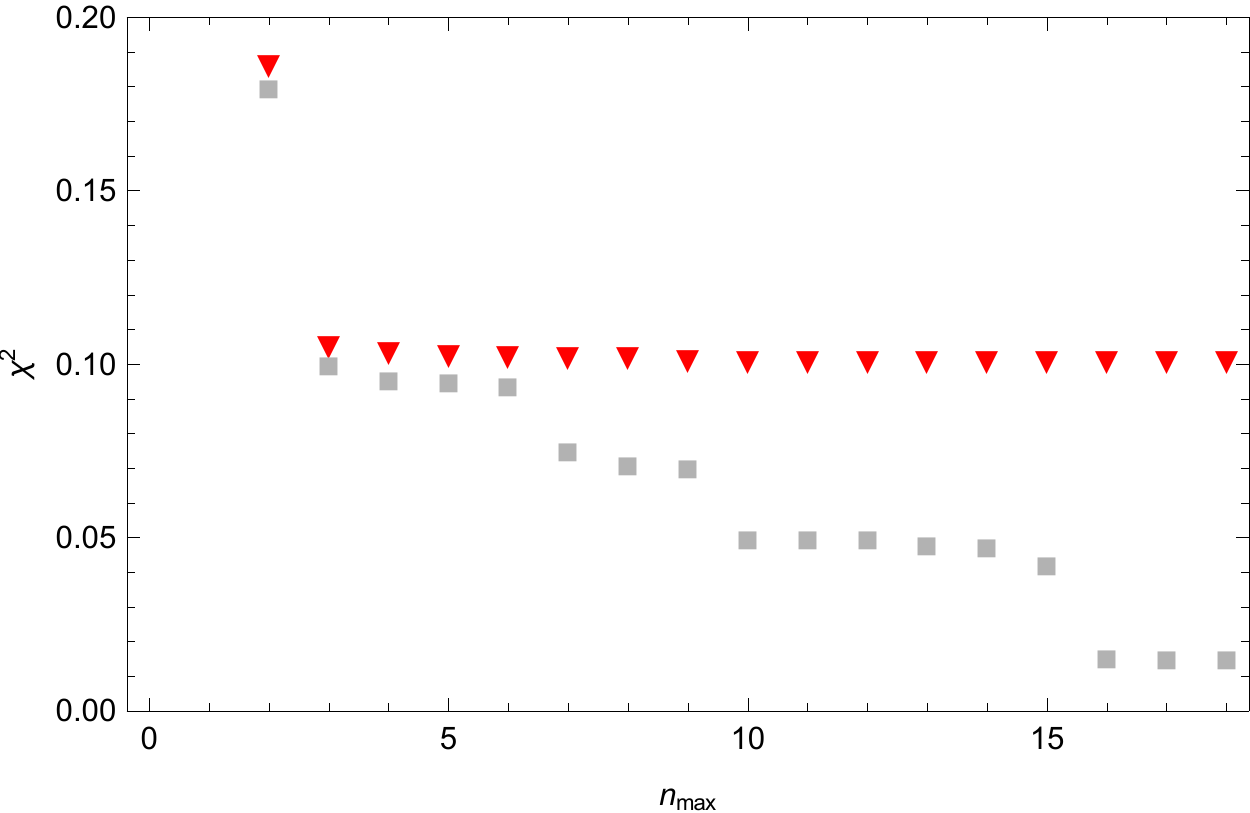}
\hspace*{0.2cm}
\includegraphics[width=0.48\textwidth]{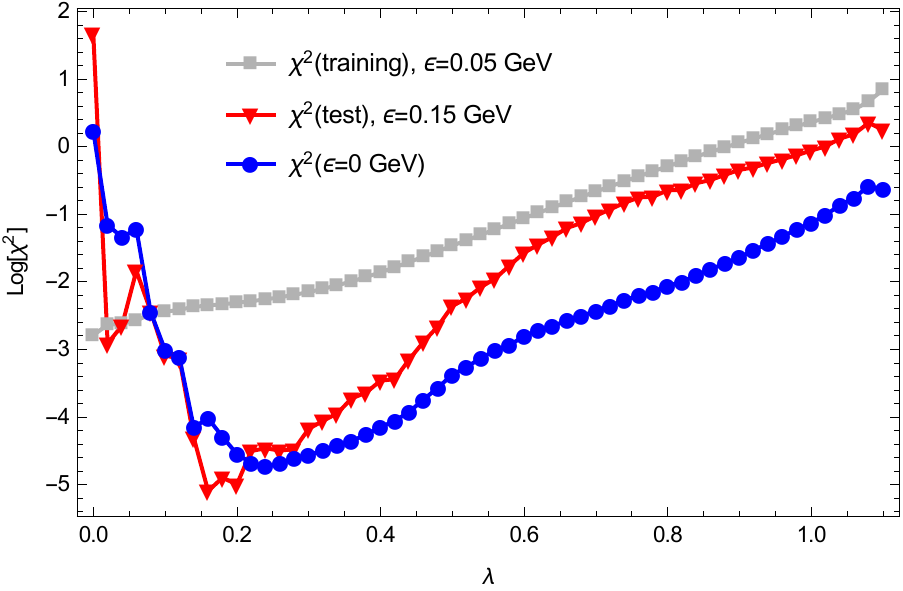}
\end{center}
\caption{ 
{\bf Left}: The $\chi^2$ as a function of the degree of the fit polynomial,
$n_{\rm max}$. While the $\chi^2$ of the unconstrained fits (gray squares)
monotonically decreases, a finite penalty factor of $\lambda=\hat\lambda_{\rm
opt}=0.2$ for $P_2$ stabilizes the result (red triangles). {\bf Right}: Cross
validation. The $\chi^2$ of the fits to the training set according to
Eq.~(\ref{chi22}) are shown with gray squares; the $\chi^2_V$ of these fits,
evaluated for the test/validation set, are indicated with red triangles; the
$\chi_t^2$ of these fits evaluated for the (unknown) true optical potential
according to Eq.~(\ref{truechi}) are displayed with blue circles. The minimum of
the $\chi_V^2(\lambda)$ of the test/validation set (red) estimates the penalty
factor $\hat\lambda_{\rm opt}\sim 0.15-0.2$ which is very close to the truly
optimal $\lambda_{\rm opt}\sim 0.2-0.3$ (blue). The absolute and relative scales
of the different $\chi^2$'s are irrelevant.
}
\label{fig:LASSO1}
\end{figure}

\begin{figure}[t]
\begin{center}
\includegraphics[width=0.48\textwidth]{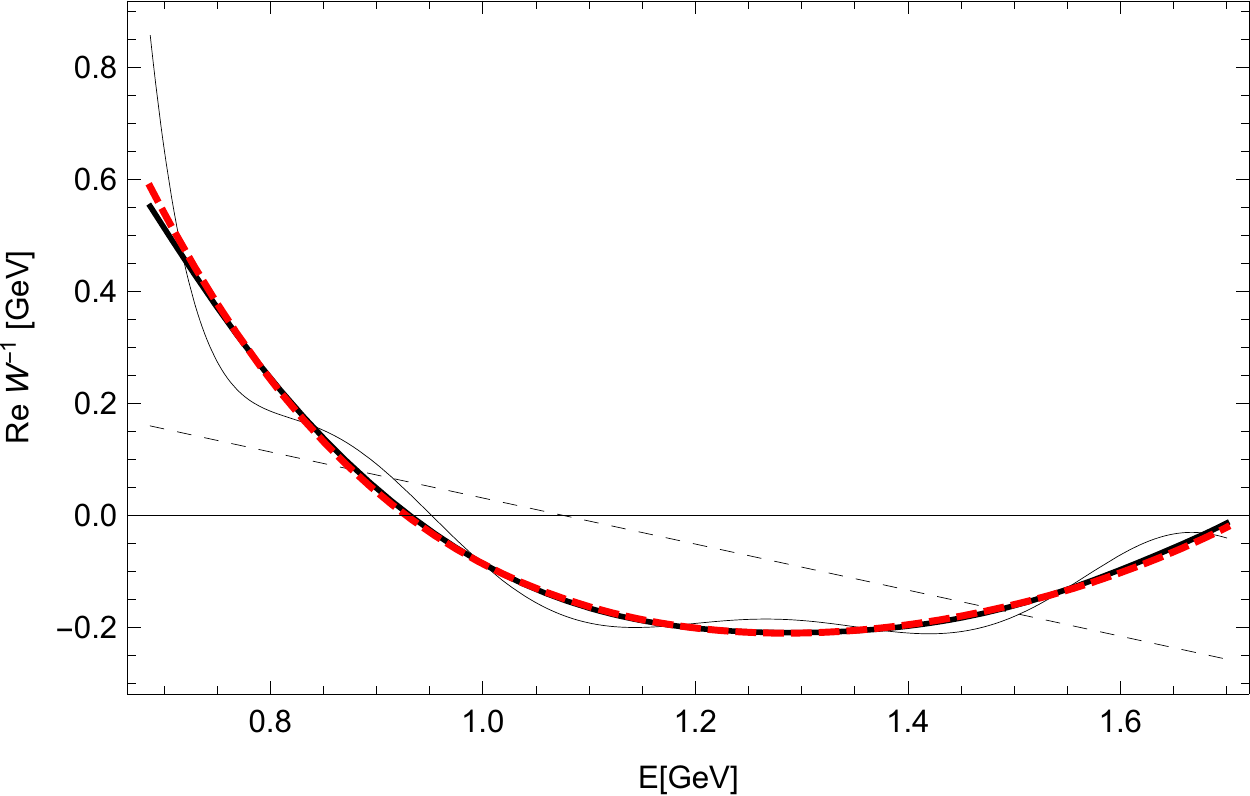}
\hspace*{0.2cm}
\includegraphics[width=0.48\textwidth]{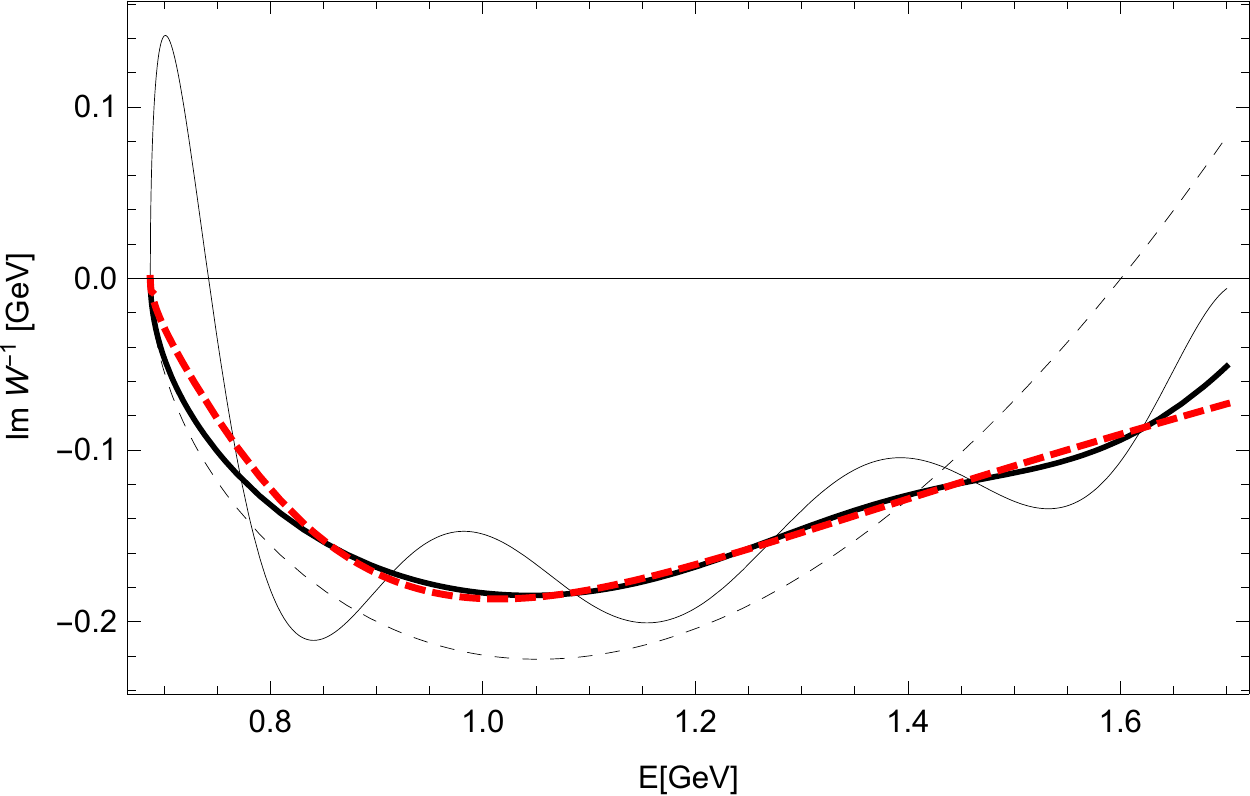}
\end{center}
\caption{Real and imaginary parts of the optical potential. The thick dashed
(red) lines show the true optical potential  $W^{-1}$. The thick solid (black)
lines show the reconstructed potential $\hat  W^{-1}$ with $\hat\lambda_{\rm
opt}=0.2$. The thin lines show a largely under-constrained result (thin solid,
oscillating lines) with $\lambda=0.05$ and a largely over-constrained result
(thin dashed lines) with $\lambda=1$.}
\label{fig:LASSO2}
\end{figure}

The basic idea of this method is to fit the optical potential $\hat
W_L^{-1}(E+i\varepsilon)$ from Eq.~\eqref{eq:realaxis} at complex energies in
the whole energy range with a suitable Ansatz.  Model selection is performed
with LASSO regularization (as explained in detail later) in combination with
cross validation. Such methods have the advantage that basic properties of the
optical potential, like Schwartz's reflection principle and threshold behavior,
can be built in explicitly. In our problem, this is particularly simple because
the only non-analyticity is given by the branch point at the $\pi\eta$
threshold. In more complex problems, additional non-analyticities like resonance
poles or complex branch points from multi-channel states~\cite{Doring:2009yv,
Ceci:2011ae} have to be included in the parameterization. Yet, all these
non-analyticities are situated on other than the first Riemann sheet. The
parametric and non-parametric methods proposed here use an extrapolation from
finite, but positive  $\varepsilon$ to $\varepsilon\to 0$, i.e., an
extrapolation performed on the first Riemann sheet that is analytic by
causality. 

A suitable yet sufficiently  general parameterization of the optical potential
is given by 
\begin{align}
\hat W^{-1}(E)=\sum_{j=0}^{n_{\rm max}}
\left[\left(a_j+i \,b_j\,p_{\pi\eta}\right)(E-E_0)^j\right]
\label{ansatz}
\end{align}
with real parameters $a_j,\,b_j$. The only non-analyticity of $\hat W^{-1}$ is
given by the cusp function $i\,p_{\pi\eta}$, evaluated at the complex energy $E$
(see Eq.~\eqref{winv}), that is therefore explicitly included in the Ansatz; the
rest is then analytic and can be expanded in a power series around a real $E_0$
chosen in the center of the considered energy region, in order to reduce
correlations among fit parameters (the actual value of $E_0$ is irrelevant).

To perform the effective infinite-volume extrapolation through smoothing, we
consider the minimization of the $\chi^2$,
\begin{align}\label{chi22}
\chi^2=\sum_{k=1}^m\frac{\left|\hat W^{-1}(E_k)-\hat W_L^{-1}(E_k)\right|^2}{\sigma_k^2}+P_i(a_j,\,b_j)\,,
\end{align}
where $P_i$ are penalty functions specified below. The absolute scale of the
$\chi^2$ is irrelevant. The quantity $\hat W_L^{-1}$ is fitted by sampling at
the complex energies $E_k=E_{\rm min}+k\,\delta E+i\varepsilon$
($\varepsilon=0.05$~GeV) over the considered energy range $E_{\rm min}\leq E\leq
E_{\rm max}$ with a step $\delta E=10$~MeV, and assigning an arbitrary error of
$\sigma_k=\sigma=1$~GeV. Note that in cross validation (to be specified later),
the position of the minimal $\chi^2$ determines the size of the penalty, i.e.,
the size of $\sigma$ is irrelevant. The infinite-volume optical potential is
then obtained by simply evaluating $\hat W^{-1}$ at real energies, i.e., setting
$\varepsilon=0$.

If we assume for the moment that the penalty function $P_i$ in Eq.~\eqref{chi22}
is zero, then it is clear that the minimized $\chi^2$ is a monotonically
decreasing function of the degree of the fit polynomial $n_{\rm max}$. This is
demonstrated by the gray squares in   Fig.~\ref{fig:LASSO1} (left panel).
Apparently, the fit stabilizes first for $n_{\rm max}=3-6$, which might lead to
the wrong conclusion that an optimal smoothing had been obtained. Then, for
higher $n_{\rm max}$, another plateau is reached at $n_{\rm max}=7-9$ and then
another one for $n_{\rm max}=10-14$. Thus, without an additional criterion, one
cannot decide which $n_{\rm max}$ is optimal.

In general, for a small $n_{\rm max}$, the smoothing will be too aggressive
(large $\chi^2$), while for too large values of $n_{\rm max}$ the fit will start
following the oscillations (Fig.~\ref{fig:oscillations}), resulting in a low
$\chi^2$ but missing the point of smearing the optical potential. These two
extreme cases are illustrated in Fig.~\ref{fig:LASSO2} with the thin dashed and
thin solid lines, respectively\footnote{These curves are derived in  a similar
but slightly different context, see below. However, they still may serve as a
good illustration for the statement given here.}. There is obviously a {\it
sweet spot} for $n_{\rm max}$. Model selection refers to the process of
determining this spot as outlined in the following. 

Model selection for the fit~(\ref{ansatz}) is formally introduced through a
penalty $P(a_j,b_j)$ imposed on the fit parameters. The penalty is formulated
using the LASSO method developed by Tibshirani in 1996~\cite{Tib0}. See also
Refs.~\cite{Tib1, Tib2} for an introduction into the topic. The LASSO method has
been recently applied in hadronic physics for the purpose of amplitude
selection~\cite{Guegan:2015mea}.

A natural choice to suppress oscillations is to penalize the modulus of the
second derivative~\cite{Tib1},
\begin{align}\label{p1}
P_1(a_j,\,b_j)=\lambda^4\,
\int\limits_{E_{\rm min}+i\varepsilon}^{E_{\rm max}+i\varepsilon} dE\,
\left|\frac{\partial^2 \hat W^{-1}(E)}{\partial E^2}\right| \,,
\end{align}
where the integral is performed along a straight line in the complex plane.
Another choice is to penalize only the polynomial part of the Ansatz
(\ref{ansatz}), i.e., removing the $p_{\pi\eta}$ factor that has an inherently
large second derivative at the $\pi\eta$ threshold,
\begin{align}\label{p2}
P_2(a_j,\,b_j)=\lambda^4\,
\int\limits_{E_{\rm min}+i\varepsilon}^{E_{\rm max}+i\varepsilon} dE\,
\left(\left|\frac{\partial^2}{\partial E^2}
\sum_{j=0}^{n_{\rm max}}
a_j(E-E_0)^j\right|+\left|\frac{\partial^2}{\partial E^2}
\sum_{j=0}^{n_{\rm max}} b_j(E-E_0)^j\right|\right) \,.
\end{align}
Including $\lambda$ to the fourth power is done in order to have a clearer
graphical representation of the penalty factor in subsequent plots. Imposing a
penalty, the decrease of $\chi^2$ with $n_{\rm max}$ is eventually stabilized,
as shown by the red triangles in Fig.~\ref{fig:LASSO1} (left panel) for some yet
to be determined value of $\lambda$. Clearly, the minimized $\chi^2$ from
Eq.~\eqref{chi22} is a monotonically increasing function of $\lambda$ as
demonstrated by the gray squares in Fig.~\ref{fig:LASSO1} (right panel) for the
penalty function $P_2$.

The fitted data ($\varepsilon=0.05$~GeV) form the so-called {\it training
set}~\cite{Tib0}. The main idea of cross validation to determine the sweet spot
of $\lambda$ is as follows (for more formal definitions and $k$-fold cross
validation, see Refs.~\cite{Tib0,Tib1,Tib2}): after a random division of a given
data set into {\it training} and {\it test/validation} sets, the fit obtained
from the training set is used to evaluate its $\chi^2$ with respect to the
test/validation set, called $\chi^2_V$ in the following (without changing fit
parameters and setting $P_i=0$). For too large values of $\lambda$, both values
of $\chi^2$ from training and from test/validation sets will be large. For too
small $\lambda$, the fit to the training set is too unconstrained and sensitive
to unwanted random properties such as fluctuations in the training data.
However, those unwanted random properties are different in the validation set,
leading to a {\it worse} $\chi^2_V$ for too small $\lambda$. It is then clear
that $\chi^2_ V(\lambda)$ exhibits a minimum at the sweet spot
$\lambda=\hat\lambda_{\rm opt}$. 

Here, we cannot meaningfully divide the data set randomly. Instead, we have to
look for data, for which the physical property (infinite-volume optical
potential) is unchanged, but the unphysical property (oscillations from
finite-volume poles) is changed. This is naturally given by $\hat W_L^{-1}$ but
evaluated for a substantially different value of $\varepsilon$ (we choose
$\varepsilon=0.15$~GeV). The analytic form of Eq.~(\ref{ansatz}) ensures
that the infinite-volume optical potential can be analytically continued to
different values of $\varepsilon$, and only the unwanted finite-volume
oscillations are different for different $\varepsilon$. Indeed, as indicated
with the red triangles in Fig.~\ref{fig:LASSO1} (right panel), $\chi^2_V$
exhibits a clear minimum at $\lambda=\hat\lambda_{\rm opt}\sim 0.2$. The
potential dependence of the this value on the chosen $\varepsilon$ is discussed
below.

Furthermore, in this example, we know the underlying optical potential and can
simply determine the (generally unknown) truly optimal value for $\lambda$,
$\lambda_{\rm opt}$ by evaluating the $\chi^2$ of the estimate of the optical
potential, $\hat W^{-1}$, with respect to the true optical potential on the real
axis, $W^{-1}$,
\begin{align}\label{truechi}
\chi^2_t(\lambda)=\sum_{k=1}^m\frac{\left|\hat W^{-1}({\rm Re}\,E_k)
-W^{-1}({\rm Re}\,E_k)\right|^2}{\sigma^2} \,.
\end{align}
Note that the quantity $\chi^2_t(\lambda)$ (implicitly) depends on $\lambda$,
because the quantity $\hat W^{-1}({\rm Re}\,E_k)$ was determined at a fixed
value of $\lambda$. The quantity $\chi^2_t$ is shown with the blue filled
circles  in Fig.~\ref{fig:LASSO1} (right panel). Its minimum at
$\lambda=\lambda_{\rm opt}$ is very close to the minimum of the validation
$\chi^2_V$ at $\lambda=\hat\lambda_{\rm opt}$, demonstrating that cross
validation~\cite{Tib1} is indeed capable of estimating the optimal penalty in
our case.  

Instead of using the penalty function $P_2$, one can also choose $P_1$, see
Eqs.~\eqref{p1} and \eqref{p2}. The estimated $\hat \lambda_{\rm opt}$ given by
the minimum of $\chi^2_V$ will, of course, change. But, again, it was checked
that the new $\hat \lambda_{\rm opt}$ is very close to the new $\lambda_{\rm
opt}$ given by the minimum of $\chi^2_t$. Similarly, we have checked other forms
of penalization, with the same findings: imposing penalty on the third
derivative, variation of the  value of $\varepsilon$ for the training set, and
variation of the value of $\varepsilon$ for the test/validation set. The only
restriction is that the $\varepsilon$ of the test/validation set has to be
chosen sufficiently larger than $\varepsilon$ of the training set for a minimum
in $\chi^2_V$ to emerge --- if the two $\varepsilon$'s are too similar, the
oscillations are too similar and no minimum in $\chi^2_V$ is obtained. Also,
$n_{\max}$ has to be chosen high enough so that, at a given $\varepsilon$ for
the training  set, the fit is capable of fitting oscillations (for small
$\lambda$) which is a prerequisite for a minimum in $\chi^2_V$ to appear. In all
simulations we have chosen $n_{\rm max}=18$ although $n_{\rm max}\sim 7$ would
suffice as the left panel of Fig.~\ref{fig:LASSO1} shows.

For the initially considered case, using $P_2$ for the penalty,
$\varepsilon=0.05$~GeV for the training set, and $\varepsilon=0.15$~GeV
for the test/validation set, the resulting optical potential is shown in
Fig.~\ref{fig:LASSO2} with the thick black solid lines. For comparison, the true
optical potential is shown with the thick red (dashed) lines. The optical
potential is well reconstructed over the entire energy range. At the $\pi\eta$
threshold, the reconstructed potential reproduces the square-root behavior due
to the explicit factor $p_{\pi\eta}$ in the parameterization~(\ref{ansatz}). The
reconstructed potential explicitly fulfills Schwartz's reflection principle and
its imaginary part is zero below threshold. At the highest energies, small
oscillations become visible originating from the upper limit of the fitted
region at $E_{\rm max}=1.7$~GeV. Here, the smoothing algorithm, that is an
averaging in energy, has simply no information on $\hat W_L^{-1}$ beyond $E_{\rm
max}$. Note that in the numerical  simulation of the next section, that uses
re-sampling techniques and realistic error bars, these small oscillations
themselves average out over the Monte-Carlo ensemble, simply resulting in a
widened, but smooth, error band at the highest energies.

For illustration, we also show in Fig.~\ref{fig:LASSO2} a largely
under-constrained result (too small $\lambda$, thin solid lines) in which the
oscillations from the finite-volume poles in $\hat W_L^{-1}$ survive. The
opposite case, i.e., an over-constrained fit with too large $\lambda$, is shown
with the thin dashed lines exhibiting too large of a penalization on the second
derivative.


\subsubsection*{Non-parametric method}

The advantage of non-parametric methods lies in its blindness of analytic
structures, which, however, also leads to the fact that threshold behavior and
Schwartz' reflection principle cannot be implemented easily. As a particular
method, we utilize an approach, commonly used in image processing applications.
This approach goes under the name of Gaussian smearing. The basic idea of the
Gaussian smearing is quite simple: for a given set of uniformly distributed
data, any data point is replaced by a linear combination of its neighboring data
points (within a given radius $r$), with individual weights, $w(x)$ given by
\begin{align}\label{eq:GAUSS}
w(x)\propto e^{-\frac{x^2}{2\sigma_0^2}}\,.
\end{align}
Here, $x$ and $\sigma_0$ denote the distance of the individual points from the
central one and the standard deviation. Typically, the latter value is chosen to
match the radius of the smearing by $\sigma_0=r/2$. Therefore, the only
undetermined quantity is given by the smearing radius $r$.

The general prescription to determine the smearing radius should rely on the
properties of the original data only. Recall that the latter is determined by
the function $\hat W_L^{-1}$ in Eq.\eqref{eq:realaxis}, which splits up into a
real and an imaginary set, when evaluated at the energy $E+i\varepsilon$ for a
fixed $\varepsilon>0$ and uniformly distributed values of $E$. Therefore, after
the fits to the (synthetic) lattice data are performed, the scale of the
structures to be smeared is determined by the distance between two poles, see
Fig.~\ref{fig:twisted-fits}. Of course, since the poles are not distributed
uniformly over the whole energy range, one could argue in favor of using
different values of $r$ for different energies. It is also clear that constraint
on the standard deviation $\sigma_0=r/2$ affects the result of the smoothing.
However, in order not to over-complicate the procedure, in the following we
choose the smearing radius to be twice as large as the typical (average)
distance between two  poles. If the radius is much larger than this, physical
information (i.e. the functional form of the optical potential) will be smeared
out. If, however, the radius is much smaller than this value, then the
(unphysical) oscillations will remain, preventing the reconstruction of the
underlying optical potential. The situation is in fact very similar to the
under- and over-constrained results, discussed in the previous section for the
too small and too large values of $\lambda$.

After the parameters of the smearing kernel \eqref{eq:GAUSS} are fixed, the
method is applied to the sets of real and imaginary parts of $\hat W_L^{-1}$ at
a fixed $\varepsilon>0$. Then the procedure is repeated, each time assuming
slightly smaller value of $\varepsilon$ than before. In the final step, a simple
(polynomial) extrapolation is performed to real energies, i.e.
$\varepsilon=0$, to obtain the final result of this procedure, namely $\hat
W^{-1}(E)$.

\medskip
In this section, we have demonstrated that the real and imaginary parts of the
optical potential can be reconstructed from the pseudophase measured on the
lattice for real energies, $W_L^{-1}$, if the analytic continuation into the
complex plane is performed. Two distinct methods are presented to smear the
oscillations which emerge from the analytic continuation, and to recover the
optical potential for real energies. It remains to be seen, how the pseudophase
can be measured in practice. This issue will be considered in the
Sect.~\ref{sec:realisticpseudophase} where a realistic numerical simulation will
be carried out as well.


\section{Reconstruction of the optical potential}
\label{sec:realisticpseudophase}

The quantity $W_L^{-1}(E)$, which is used to extract the optical potential,
along with the energy $E$, depends on other external parameters, say, on
the box size $L$, boundary conditions, etc. In the fit to $W_L^{-1}(E)$, the
values of these parameters have to be fixed. Otherwise, for example, the
position of the poles in $W_L^{-1}(E)$ will be volume-dependent and a fit is not
possible. Hence, we are quite restricted in the ability to scan the variable
$E$: the knob, which tunes $E$, must leave all other parameters in the
pseudophase intact.

\begin{figure}[t]
\begin{center}
\includegraphics[width=\linewidth]{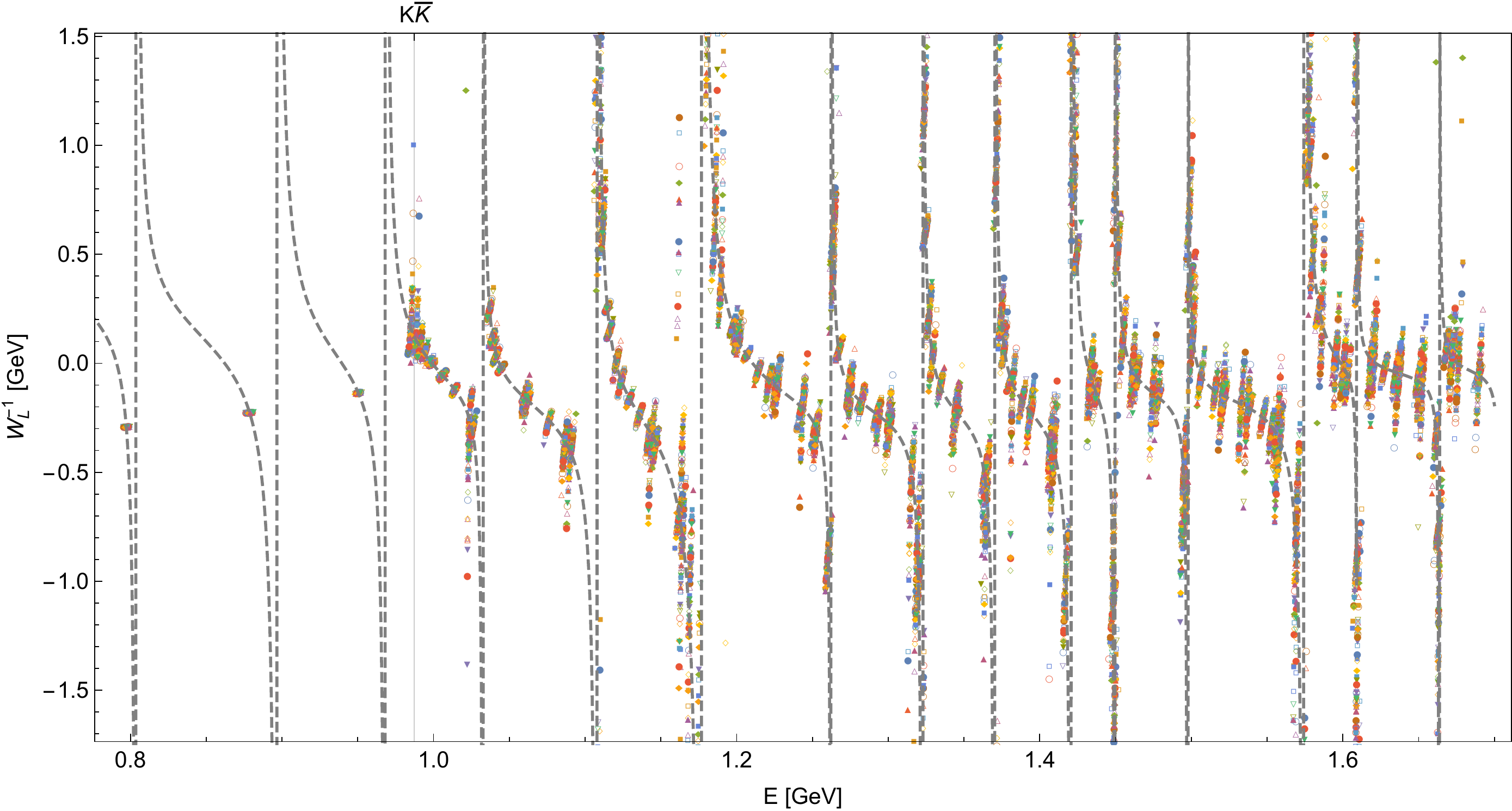}
\end{center}
\caption{
Subset (75 sets) of the re-sampled lattice data, where each type of marker
symbols shows the set of $189$ energy eigenvalues, randomly distributed with
$\Delta E=1$~MeV around the central energy eigenvalues, extracted from
Eq.~\eqref{eq:WL} imposing twisted boundary condition. The gray dashed line
shows the actual amplitude $W_L^{-1}(E)$ to guide the eye.}
\label{fig:twisted-sets}
\end{figure}

\begin{figure}[t]
\begin{center}
\includegraphics[width=\linewidth,trim= 0.1cm 3cm 0.1cm 3cm]
{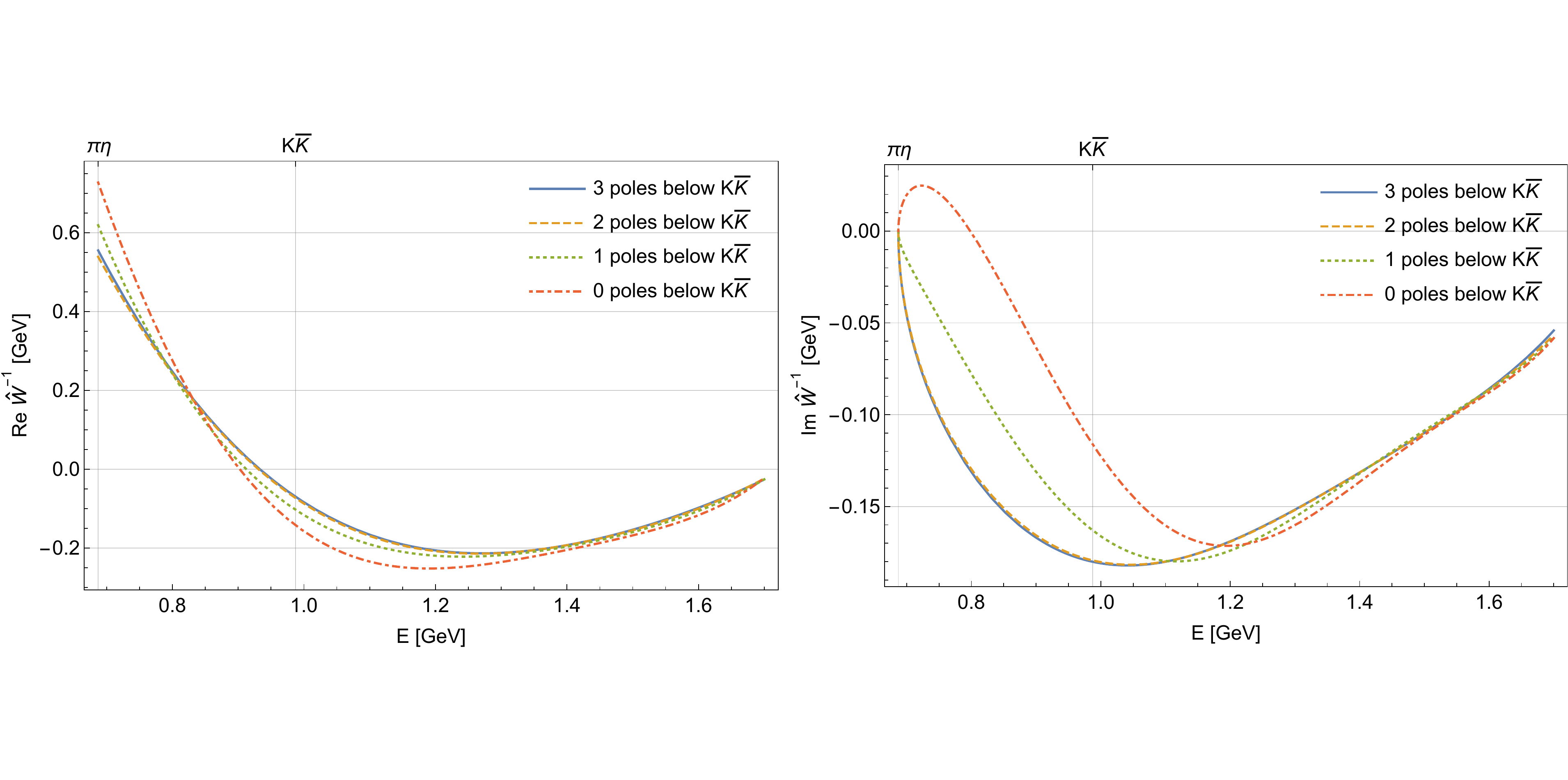}
\end{center}
\caption{Comparison of different scenarios with respect to the number of poles
reconstructed below the primary threshold.  The curves were produced by using
the parameters of the perfect fit from the Sect.~\ref{sec:complex_plane},
but neglecting a certain number of poles below the $K\bar K$ threshold.}
\label{fig:doweneedpoles}
\end{figure}

\begin{figure}[t]
\begin{center}
\includegraphics[width=\linewidth]{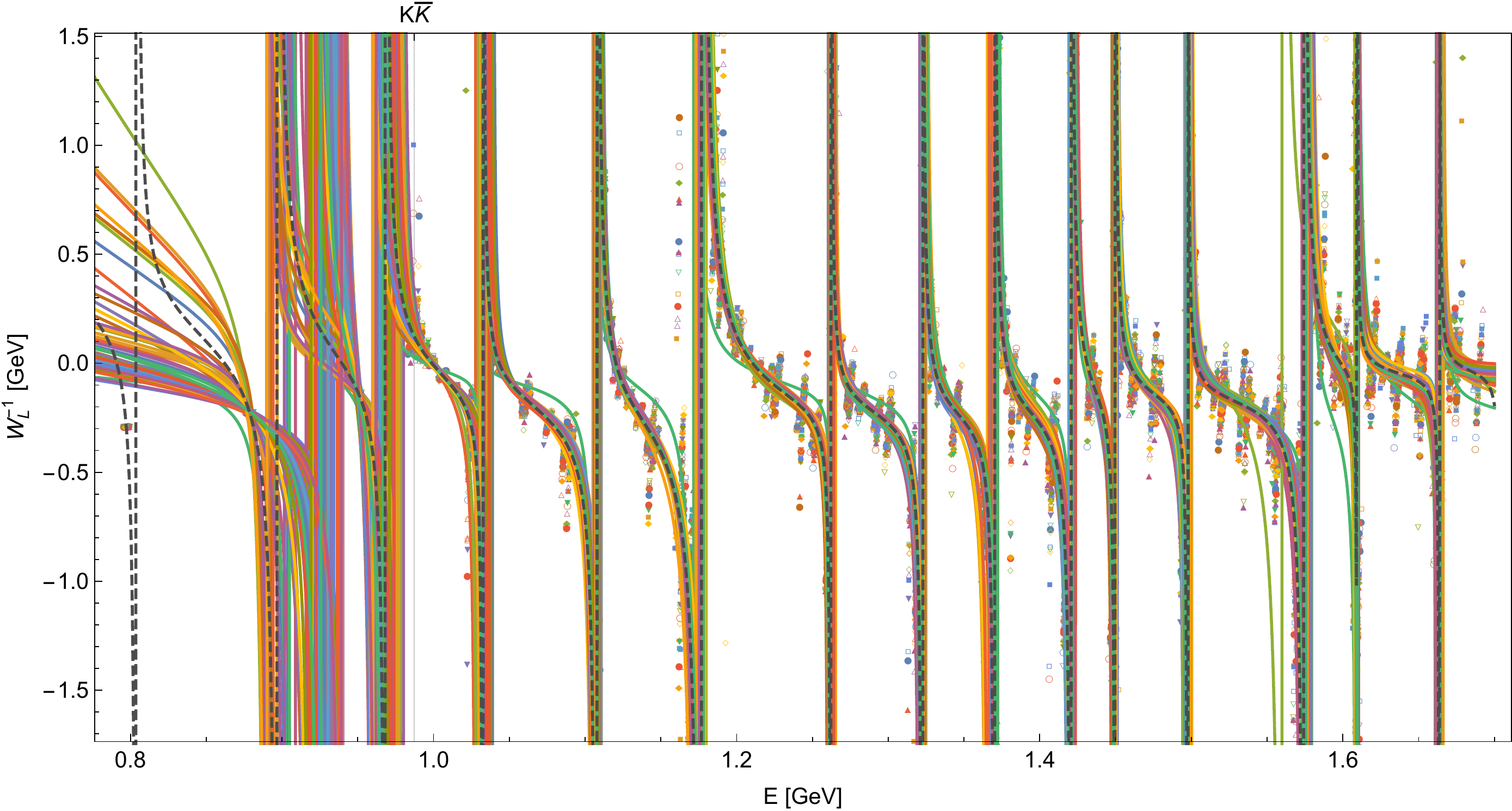}
\end{center}
\caption{A subset (75 sets) of the fits of Eq.~\eqref{eq:WL} to the
synthetic lattice data as described in the main text. Different curves represent
fits to different sets of re-sampled synthetic lattice data corresponding to the notation
of Fig.~\ref{fig:twisted-sets}. The gray dashed line shows the
actual amplitude $W_L^{-1}(E)$ to guide the eye.}
\label{fig:twisted-fits}
\end{figure}

\begin{figure}[t]
\begin{center}
\includegraphics[width=\linewidth]{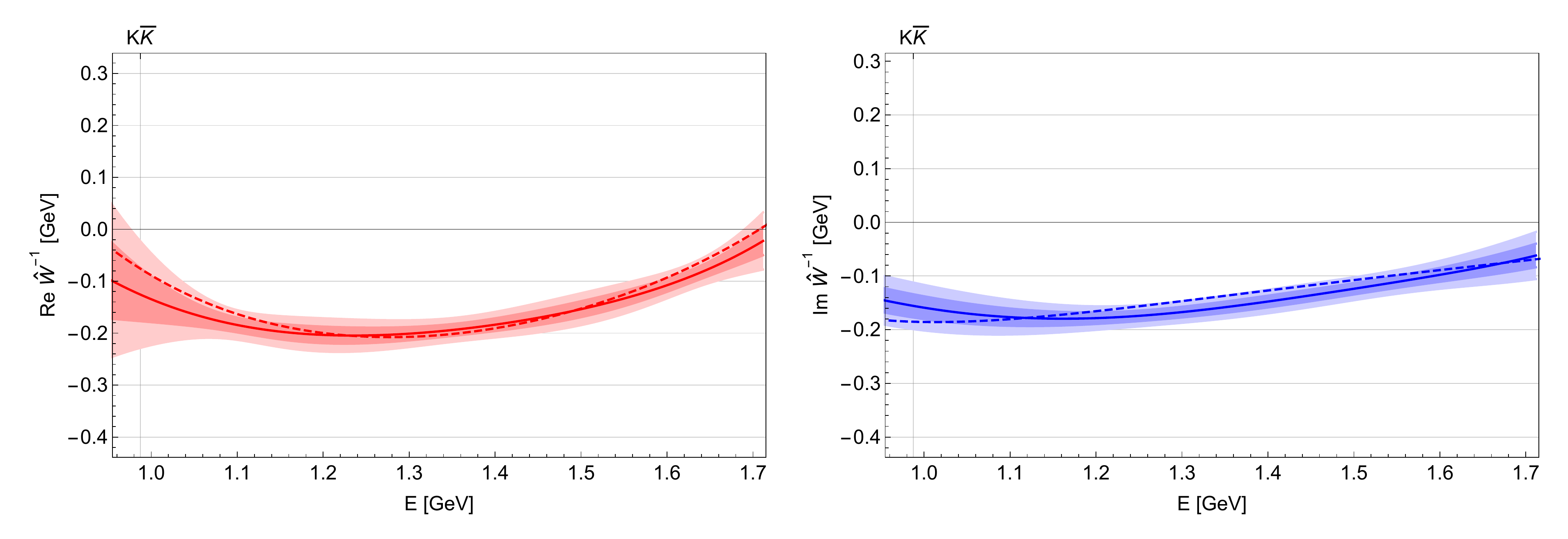}
\includegraphics[width=\linewidth]{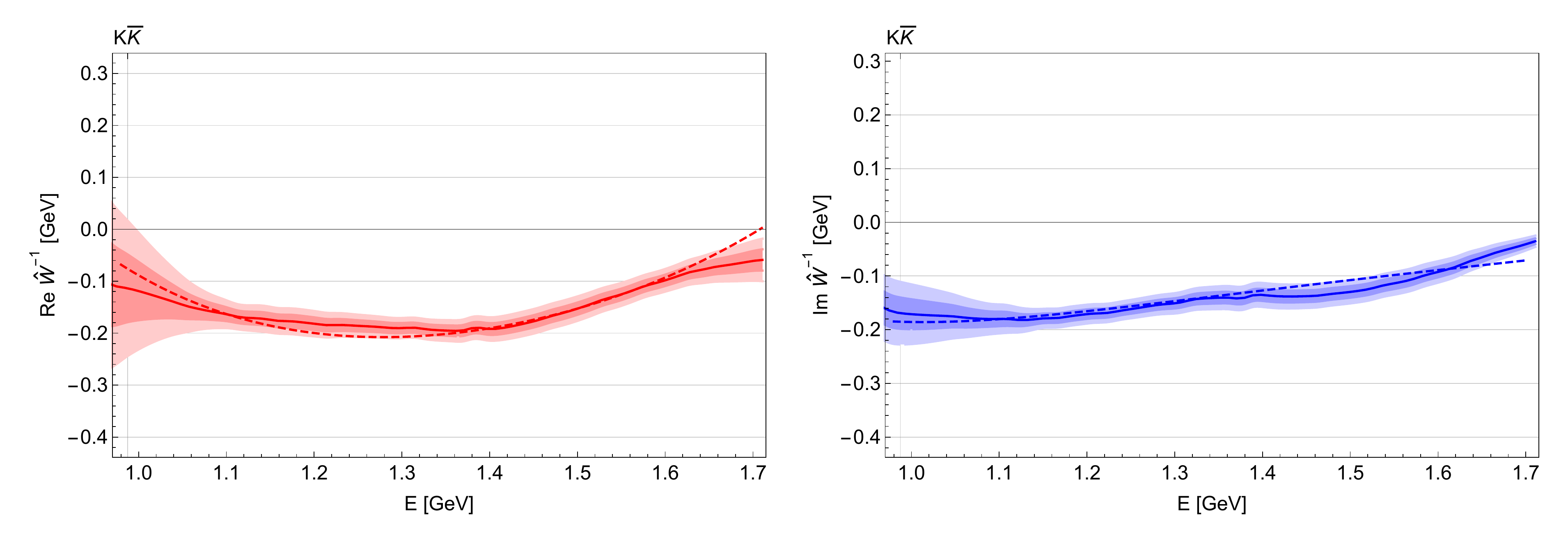}
\end{center}
\caption{Results of the smearing and extrapolation to real energies using
parametric method (top) and Gaussian smearing (bottom). The full lines show the
 average of the re-sampling of all sets, whereas the darker (lighter)
bands show the corresponding 1 (2) $\sigma$ error bands. The exact infinite
volume solution is shown by the dashed lines for comparison.}
\label{fig:final1MeV}
\end{figure}

\begin{figure}[t]
\begin{center}
\includegraphics[width=\linewidth]{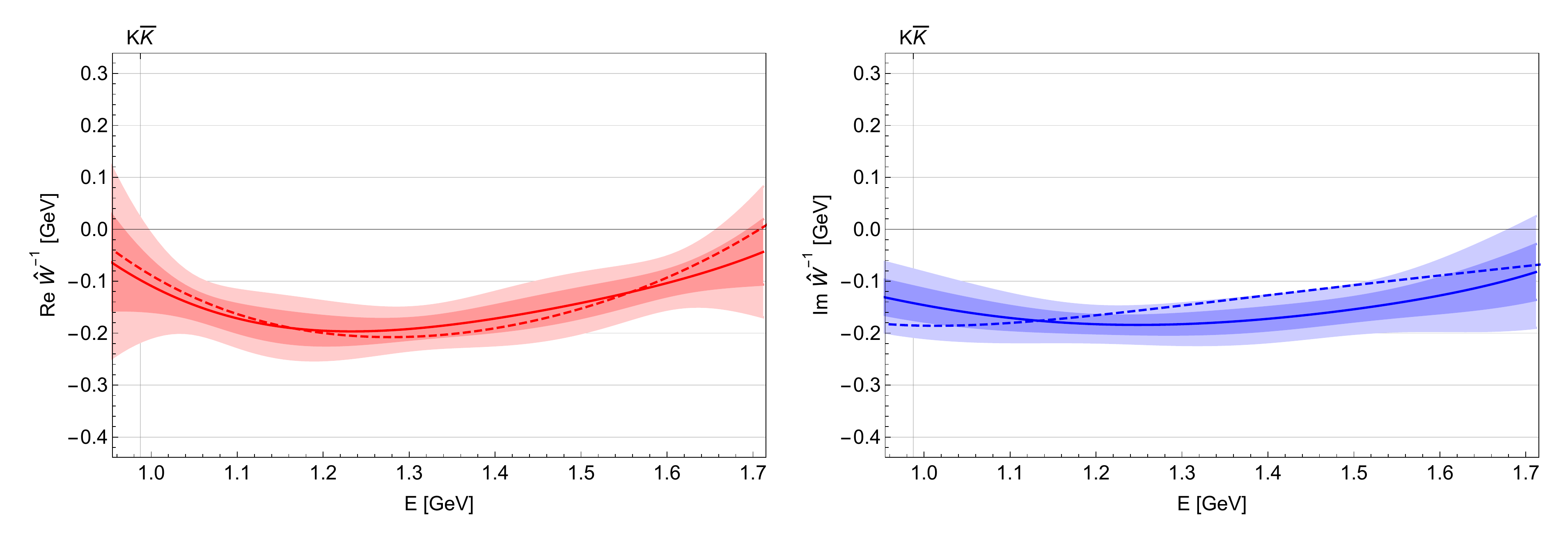}
\includegraphics[width=\linewidth]{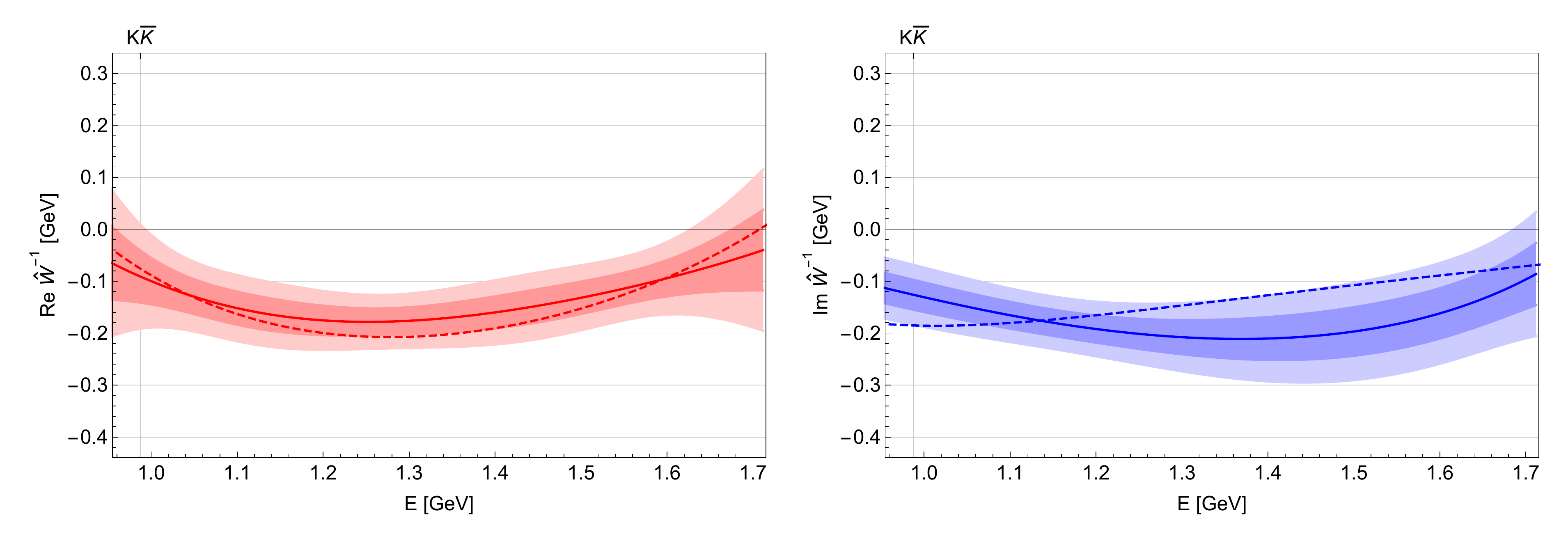}
\end{center}
\caption{Results of the smearing and extrapolation to real energies using
parametric method for synthetic lattice data with $\Delta E=2$~MeV (top) and
$\Delta E=3$~MeV (bottom). The full lines show the average of the
re-sampling of all sets, whereas the darker (lighter) bands show the
corresponding 1 (2) $\sigma$ error bands. The exact infinite volume solution is
shown by the dashed lines for comparison.}
\label{fig:final23MeV}
\end{figure}


\subsection{Partially twisted boundary conditions}\label{sec:partialtwisting}

In certain systems, there indeed exists a possibility to scan the energy within
a given range in this manner. It is provided by the use of twisted boundary
conditions and  can be realized, e.g., in the coupled-channel $\pi\eta-K\bar K$
scattering. Namely, as was discussed in
Refs.~\cite{Lage-scalar,Agadjanov-twisted}, in this system it is possible to
apply (partially) twisted boundary conditions so that, when the twisting angle
is changed continuously, the $K\bar K$ threshold moves, whereas the $\pi\eta$ 
threshold stays intact. This can be achieved, for example, by twisting the light
$u,d$ quarks by the same angle and leaving the $s$-quark with periodic boundary
conditions.  This will lead to the modification of the secular
equation~\eqref{eq:secular}, replacing $Z_{00}(1;q_{K\bar K}^2)$ by
\begin{equation} Z_{00}^\theta(1;q_{K\bar
K}^2)=\frac{1}{\sqrt{4\pi}}\,\sum_{{\bf n}\in{\mathbb Z}^3}\frac{1}{\left({\bf
n}+\boldsymbol{\theta}/2\pi\right)^2-q_{K\bar K}^2}\, . \end{equation} 
The expression for $W_L^{-1}(E)$ remains the same and does not contain the
twisting angle $\boldsymbol{\theta}$.

The method can be used to study the isospin $I=1$ scattering in the
$\pi\eta-K\bar K$ system. As shown in Ref.~\cite{Agadjanov-twisted}, despite the
presence of the annihilation diagrams, the partial twisting in this case is
equivalent to the full twisting, if the light quarks are twisted, whereas
twisting of the $s$-quark does not lead to an observable effect. As a rule of
thumb, one expects that the partial twisting of a given quark will be equivalent
to full twisting, only if this quark line goes through the diagram without being
annihilated (of course, a rigorous proof of this statement should follow by
using effective field theory methods~\cite{Agadjanov-twisted}). In our case, we
could choose to work with the state with maximal projection of the isospin, say
$I=1,I_3=1$. This state contains one $u$-quark and one $\bar d$-quark, which
cannot be annihilated. Choosing the same twisting angle for both quarks, the
system stays in the center-of-mass frame and the pseudophase becomes independent
from the  twisting angle, as required. From the above discussion it is also
clear that using our method for the extraction of the optical potential in the
channel with isospin $I=0$ implies the use of full twisting instead
of partial twisting.

The same trick can be used to study the $Z_c(3900)$ and $Z_c(4025)$ states,
which both have isospin $I=1$. Twisting $u$- and $d$-quarks by the same angle,
the $D$- and $D^*$-mesons will get additional momenta proportional to the
twisting angle, whereas the $J/\psi$, $h_c$ and $\pi$-mesons will not.
Consequently, one may choose the channels containing the $D$ and $D^*$ mesons as
the primary ones (in our nomenclature) and regard every other channel as
secondary. For this choice, the pseudophase will not depend on the twisting
angle.

Last but not least, an unconventional twisting procedure was used in the study
of the $J/\psi\phi$ scattering from $Y(4140)$ decays~\cite{Sasaki}. Namely,
in that work the $c$- and $s$-quarks were twisted by the angles
$\boldsymbol\theta$ and $-\boldsymbol\theta$, respectively, whereas their
Hermitean conjugates $\bar c$, $\bar s$ were subject to periodic boundary
conditions. Albeit in the particular case of $J/\psi\phi$ scattering the
twisting cannot be used for the extraction of the optical potential, one could
not exclude a possibility that this kind of twisting could be applied in other
systems for this purpose. For this reason, we consider in detail this case of
(unconventional) twisting in  App.~\ref{app:twisting}.


\subsection{Analysis of synthetic data}\label{sec:simulation}

In the following, we shall reconstruct the optical potential from a synthetic
lattice data set generated by the chiral unitary approach of
Ref.~\cite{Oller}. Twisted boundary conditions are applied as described
above, and the box size is taken to be $L=5M_\pi^{-1}$. In the first stage of
our analysis we have observed that more than 100 energy eigenvalues are required
to extract the potential in the considered, and quite wide, energy range from $E=2M_K$ to $E=1.7$~GeV. To produce the synthetic data, we consider the following set
of six different twisting angles
\begin{align}
\boldsymbol{\theta}=
\begin{pmatrix}
0\\0\\0
\end{pmatrix},~\begin{pmatrix}
0\\ 0\\ \pi
\end{pmatrix},~\begin{pmatrix}
0\\ \pi\\ \pi
\end{pmatrix},~\begin{pmatrix}
\pi\\ \pi\\ \pi
\end{pmatrix},~\begin{pmatrix}
0\\ 0\\ \pi/2
\end{pmatrix},~\begin{pmatrix}
0\\ \pi/2\\ \pi/2
\end{pmatrix}.
\label{thetas}
\end{align}
For these values, $Z_{00}^\theta(1;q_{K\bar K}^2)$ has the smallest
number of poles. This requirement is important, when the energy eigenvalues are
measured with a finite accuracy. Then, in proximity of its poles, the function
$Z_{00}^\theta(1;q_{K\bar K}^2)$ will exhibit a very large
uncertainty. Solving Eq.~\eqref{eq:secular} with $Z_{00}(1;q_{K\bar K}^2)$
replaced by $Z_{00}^\theta(1;q_{K\bar K}^2)$ for each of the aforementioned
angles we were able to extract $186$ energy eigenvalues above and $3$ below the
$K\bar K$ threshold. Further, in any realistic lattice simulation, the
eigenvalues will be known only up to a finite precision. To check the
feasibility of the proposed method, it is important to account for this error,
$\Delta E$, and to see how this uncertainty\footnote{Since higher excited levels
are harder to measure, the uncertainty will presumably increase with the energy.
However, in this first study we will assume constant values for $\Delta E$.} is
reflected in the final result as studied with re-sampling techniques in the
following. Therefore, we start from a sufficiently large number ($\sim 1000$) of
re-sampled lattice data sets, normally distributed around the
($189$) synthetic eigenvalues with a standard deviation of $\Delta E$. An example of 75
synthetic lattice  data sets with $\Delta  E=1$~MeV  is presented in
Fig.~\ref{fig:twisted-sets}.

In the next step, we determine the parameters of Eq.~\eqref{eq:realaxis} for
each of these sets. Prior to doing so, we have to clarify several questions:
\begin{itemize}
\item \textbf{Range of applicability.} 
Below the $K\bar K$ threshold, the function $Z_{00}^\theta(1;q_{K\bar K}^2)$
does not depend on $\boldsymbol{\theta}$ up to exponentially suppressed
contributions. Therefore, only a limited number of energy eigenvalues can be
determined. A reliable extraction of positions and residua of all four lowest
poles is not  possible because the twisting cannot generate the necessary scan
of $W_L^{-1}$ in this energy region. This means that, on the one hand, this
approach does not allow one to extract the  optical potential below the primary
($K\bar K$) threshold. On the other hand, it is crucial to recall that, due to
smearing applied in the complex energy plane, this failure will yield the wrong
real and especially imaginary parts of the reconstructed $\hat W^{-1}(E)$. This
is demonstrated in Fig.~\ref{fig:doweneedpoles}, which was produced by using the
test parameters of the perfect fit from the last section, but neglecting a
certain number of poles below the $K\bar K$ threshold. It is seen that the
imaginary part of $\hat W^{-1}$ at the primary threshold deviates by about
$50\%$, if no poles are considered below this threshold. However, already the
inclusion of the first pole below the primary threshold improves the 
description drastically. Therefore, all poles above as well as the one below the
primary threshold should be considered in the fit to the (synthetic) lattice
data. Note also that if the secondary channels open above the primary channel,
none of these complications arise.
\item \textbf{Number of poles - starting values.} 
We found that, for sufficiently many eigenvalues and $\Delta E$ of the
order of several MeV, the number of poles above the primary threshold to be
fitted can be determined, searching for a rapid sign change of
$Z_{00}^\theta(1;q_{K\bar K}^2)$. The corresponding energy eigenvalues serve us
as limits on the pole positions, while the residua are allowed to vary freely. 
\item \textbf{Highest order of the polynomial part.} 
In principle, the order of the polynomial part of
Eq.~(\ref{eq:realaxis}) is not restricted a priori. We have tested explicitly
that adding terms of fourth or fifth order in energy to the fit function yields
only a small change of the reconstructed potential. This part may be further
formalized by conducting combined $\chi^2$- and $F$-tests on the $\chi^2$
defined below.
\item \textbf{Definition of $\chi^2$.} 
The uncertainty of the (synthetic) lattice data is given by $\Delta E$ only.
Therefore, a proper definition of $\chi^2_{\rm d.o.f.}$ should account for the
difference between the measured $\{E_i|i=1,...,N\}$ and fitted eigenvalues
$\{E^{f}_i|i=1,...,N\}$ compared to $\Delta E$ for all $N$ data points. The
$E_i^f$ eigenvalues are defined as the solutions of the following equation
\begin{align}
\frac{2}{\sqrt{\pi}L}\,Z_{00}^\theta(1;q^2_{K\bar K}(E))
=\sum_j\frac{Z_j}{E-Y_j}+D_0+D_1E+D_2E^2+D_3E^3\,,
\end{align}
which is technically very intricate. The problem of finding such solutions can
be circumvented by expanding both sides of the latter equation in powers of
$(E_i^f-E_i)$ around $E_i$ for each $i=1,...,N$. Up to next-to-leading order in
this expansion, the correct quantity to minimize reads 
\begin{align}\label{chi2dof}
\chi^2_{\rm d.o.f.} = \frac{1}{N-n}\sum_{i=1}^N \frac{1}{\Delta E^2}
\left(\frac{\hat W_L^{-1}(E)-Z_{00}^{\theta_i}(1;q^2_{K\bar K}(E))}
{\left(Z_{00}^{\theta_i}(1;q^2_{K\bar K}(E)\right)'-\left(\hat W_L^{-1}(E)
\right)'}\right)^2_{E=E_i} \,,
\end{align}
where $n$ is the number of free parameters and $\boldsymbol{\theta}_i$ is the
twisting angle corresponding to the energy eigenvalue $E_i$. Note that the
$\chi^2$ in Eq.~(\ref{chi2dof}) differs from the usual definition by a
correction factor in the denominator, given by the difference of the derivatives
of the L\"uscher and the fit function.
\end{itemize}
For every member of the data sets, each consisting of 188 energy eigenvalues
(186 above and 2 below threshold), we perform a fit, minimizing $\chi^2_{\rm
d.o.f.}$ given in Eq.~\eqref{chi2dof}.  Note that the two closest energy eigenvalues
below the $K\bar K$ threshold, which are included in the fit, are assigned a
weight factor of 6, because they are measured at every value of
$\boldsymbol{\theta}$ of Eq.~(\ref{thetas}) and do not depend on its
value up to exponentially suppressed contributions. Further, the number of free
parameters $n$ is set to $32$, consisting of 13(1) pole positions and 13(1)
residua above(below) $K\bar K$ threshold, as well as 4 parameters in the
polynomial part. The minimization is performed by using the Minuit2 (5.34.14)
library from Ref.~\cite{MINUIT}. A representative subset (75 synthetic lattice
data sets) of the results of the fits is shown in Fig.~\ref{fig:twisted-fits}.
It is seen that the data are described fairly well by all fits in a large energy
region starting above the $K\bar K$ threshold.  At and below this threshold,
there is much larger spread of the fit curves describing the data. Especially
the pole at $\sim0.9$~GeV is not fixed very precisely which is quite
natural, keeping in mind the small number of synthetic data points in this
energy region.

For each of the above fits we proceed as described in
Sect.~\ref{sec:complex_plane}. First, the function $\hat W_L^{-1}(E)$ is
evaluated at the complex energies. Second, using the Gaussian smearing as well
as the parametric method discussed in
Sect.~\ref{sec:smearing}, the real and imaginary parts of the potential are
smoothened. The penalty factor $\lambda=0.28$ (see App.~\ref{app:reallambda})
and the smearing radius $r=0.2$~GeV are used in these methods, respectively.
Finally, for every energy, we calculate the average and the standard deviation
$\sigma$. The result of this procedure is presented in Fig.~\ref{fig:final1MeV}.
It is seen that both smearing methods yield very similar results. Overall, the
exact solution (the dashed line) in the considered energy region lies within 1
or 2 sigma bands around the reconstructed potential. The error band appears to
be comfortably narrow, but becomes broader around the $K\bar K$ threshold and
$E_{\rm max}=1.7$~GeV. This effect is a natural consequence of the missing 
information outside the energy region, which influences the prediction within
the energy region via smearing during the intermediate steps of the potential
reconstruction.

Furthermore, we have repeated the whole procedure of synthetic lattice data
generation, fitting and recovering of the optical potential for higher
uncertainty  of the energy eigenvalues, $\Delta E=2$~MeV and $\Delta E =3$~MeV.
The results are presented in Fig.~\ref{fig:final23MeV} and show that the error
bars grow roughly  linearly with $\Delta E$ and that the real part of the
reconstructed amplitude remains quite stable. The imaginary part is more
sensitive to the value of $\Delta E$. Further, at even higher values of
$\Delta E\sim 10$~MeV, the fit is not reliable anymore and the imaginary part
becomes very small.


\section{Conclusions}\label{sec:concl}


\begin{itemize}
\item[i)]
In the present paper, we formulate a framework for the extraction of  the
complex-valued optical potential, which describes hadron-hadron scattering in
the presence of the inelastic channels, from the energy
spectrum of  lattice QCD. An optical potential, defined in the present article,
is obtained by using causal prescription $E\to E+i\varepsilon$ for
the  continuation into the complex energy plane. It converges to the ``true''
optical potential in the limit $L\to\infty$,  $\varepsilon\to 0$. A
demonstration of the effectiveness of the method has  been carried out by
utilizing  synthetic data.
\item[ii)]
The approach requires the precise measurement of the whole tower of the  energy
levels in a given interval. The optical
potential is then obtained  through averaging over all these levels.
\item[iii)]
Moreover, the availability of this approach critically depends on our ability to
take the lattice data at neighboring energies without changing the  interaction
parameters in the secondary channels. This can be achieved, e.g., by using
(partially) twisted boundary conditions that affects the pripary 
channel only. In the paper, we consider several systems,
where the  method can be applied. It is remarkable that some candidates for the
QCD exotica are also among these systems. 

We would like to emphasize that the use of twisted boundary
conditions is only a tool, which is used to perform a continuous energy scan of
a  certain interval. Whatever method is used to measure the dependence  of the
pseudophase on energy (all other parameters fixed), our approach,  based on the
analytic continuation into the complex plane, could be  immediately applied.
\item[iv)]
The approach could be most useful to analyze systems, in which the
inelastic  channels contain three or more particles. Whereas
direct methods based on the use of multi-particle scattering equations in a
finite volume will be necessarily  cumbersome and hard to use, nothing changes,
if our approach is applied. The reason for this is that, in case of an
intermediate state with any number  of particles, the single poles are the only
singularities in any Green's function in a finite volume. 
\end{itemize}


\section*{Acknowledgments}

We would like to thank S. Aoki, G. Schierholz and C. Urbach for helpful
discussions. Financial support by the Deutsche Forschungsgemeinschaft (CRC 110,
``Symmetries and the Emergence of Structure in QCD''), the Volkswagenstiftung
under contract no. 86260, the National Science Foundation (CAREER grant No.
1452055, PIF grant No. 1415459), GWU (startup grant), The Chinese
Academy of Sciences  (CAS) President's International Fellowship Initiative
(PIFI) (Grant No. 2015VMA076), and by the Bonn-Cologne Graduate School of
Physics and Astronomy is gratefully acknowledged.

\appendix


\section{Penalty factor for a realistic set of the synthetic data}
\label{app:reallambda}

\begin{figure}[t]
\begin{center}
\includegraphics[width=0.237\textwidth]{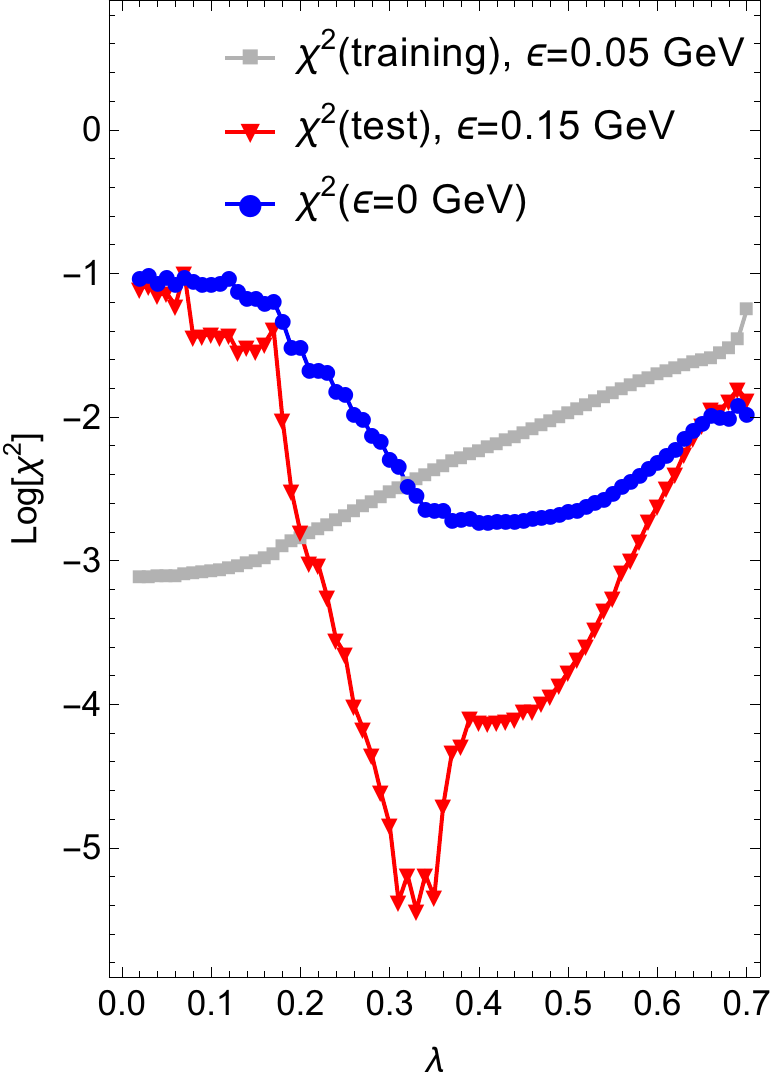}
\includegraphics[width=0.243\textwidth]
{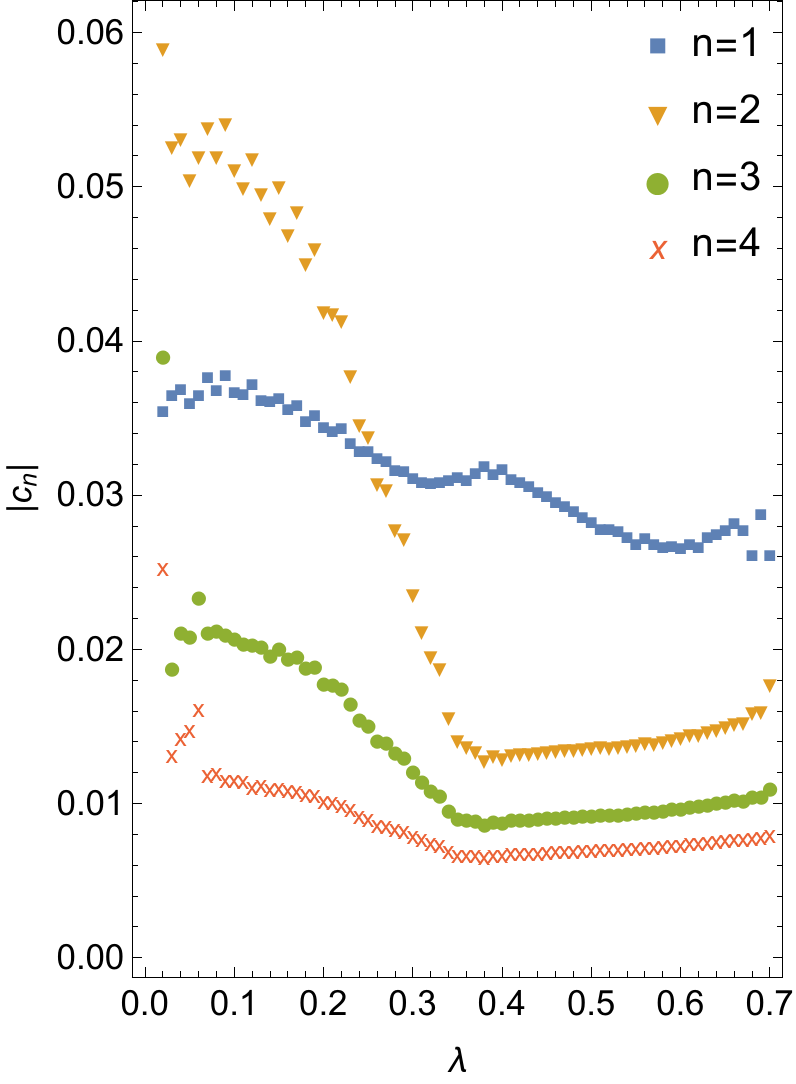}
\hspace*{0.1cm}\rule[0ex]{0.1ex}{13.5em}
\includegraphics[width=0.237\textwidth]{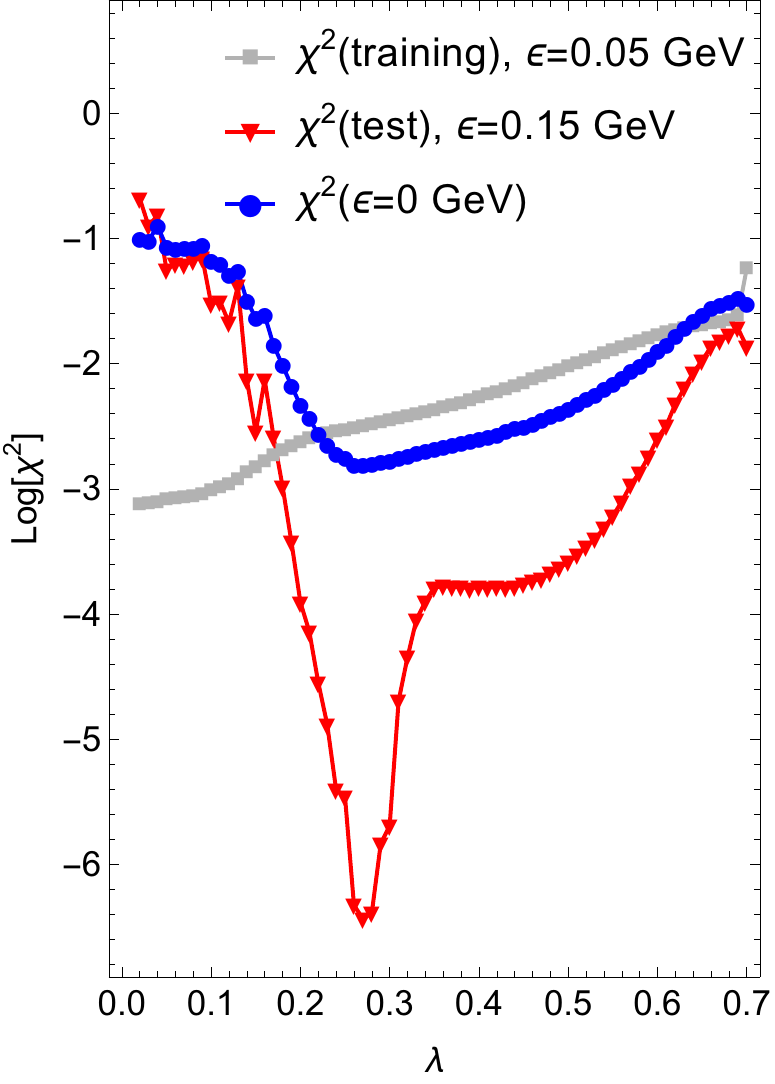}
\includegraphics[width=0.243\textwidth]
{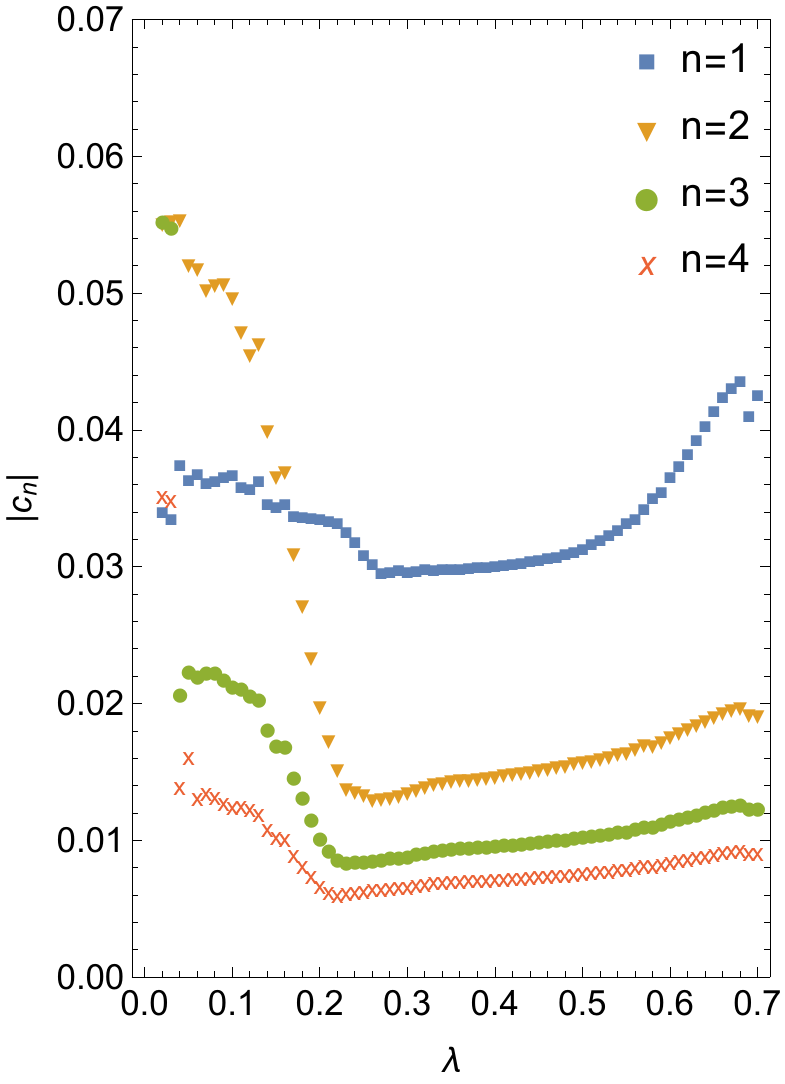}
\end{center}
\caption{Determination of $\hat\lambda_{\rm opt}$ for a realistic numerical
simulation. Notation as in Fig.~\ref{fig:LASSO1}. Left two graphs: Using the
penalization $P_1$ of Eq.~(\ref{p1}). Right two graphs: Using the penalization
$P_2$ of Eq.~(\ref{p2}). For each case, the $\chi^2$ (training set), $\chi^2_V$
(test/validation set) and $\chi^2_t$ (true $\chi^2$) are displayed.
Additionally, the moduli of the Fourier coefficients $|c_n|$, $n=1,\dots,4$ are
shown for each case. For further explanations, see text.}
\label{fig:LASSO3}
\end{figure}

In Sect.~\ref{sec:smearing},  where the parametric method for the smearing was
introduced, we assumed that the quantity $W_L^{-1}$ can be measured with no
uncertainties and at all energies from $E_{\rm min}=M_\pi+M_\eta$ to $E_{\rm
max}=1.7$~GeV. We now turn to a more realistic case, studied in the numerical
simulation in Sect.~\ref{sec:realisticpseudophase}. For this, the search for
$\hat \lambda_{\rm opt}$ is adapted to the interval from $E_{\rm min}=2M_K$ to
$E_{\rm max}=1.7$~GeV, using several $\hat W_L^{-1}$'s from the Monte-Carlo
ensemble (see description there). Fig.~\ref{fig:LASSO3} shows the $\chi^2$
behavior for the training set, the test/validation set $\chi^2_V$, and the true
$\chi^2_t$ for one arbitrarily chosen fit of the Monte-Carlo ensemble of
different $\hat W_L^{-1}$'s. Both variants of the penalty, $P_1$ and $P_2$ from
Eqs.~(\ref{p1}, \ref{p2}), are shown in the left and right panels, respectively.

As Fig.~\ref{fig:LASSO3} shows, the minima of $\chi^2_V$ (red triangles) are
even more pronounced than in the previously discussed, idealized case, leading
to $\hat\lambda_{\rm opt}=0.34$ for $P_1$ and $\hat\lambda_{\rm opt}=0.28$ for
$P_2$ (minima of the curves shown with red triangles). The minima of the
$\chi^2_t$ occur almost at the same respective values of $\lambda$ (blue filled
circles) which again demonstrates the applicability of the method. For both
penalties, we also show the moduli of the Fourier coefficients $|c_n|$,
$n=1,\dots,4$ in the respective right panels, where
\begin{equation}
c_n(\lambda)=\frac{1}{E_{\rm max}-E_{\min}}
\int\limits_{E_{\rm min}}^{E_{\rm max}} dE\, \hat W^{-1}(E)\,\,
e^{-i\,\frac{2\pi  n E}{E_{\rm max-E_{\rm min}}}} \ .
\end{equation}
Here, the infinite-volume quantity  $\hat W^{-1}(E)$ implicitly depends on
$\lambda$. These coefficients indicate the weight of the available frequencies
to built up the optical potential over a finite energy range. As long as the
potential is smooth, we expect the lowest $|c_n|$ to dominate. For decreasing
values of $\lambda$, eventually a point is reached at which the oscillations
will become noticeable and coefficients $|c_n|$ with larger $n$ will become more
relevant. Indeed, the figure shows that, close to the respective
$\hat\lambda_{\rm opt}$'s, the coefficients $|c_2|$ to $|c_4|$ exhibit a very
pronounced rise. In all simulations, which were carried out, we have observed
this behavior. This suggests that the $\lambda$-dependence of the Fourier
coefficients can be used as a tool to cross-check the results from cross
validation.

As a final remark, the value of $\hat\lambda_{\rm opt}$ itself carries
uncertainty that can be estimated by $k$-fold cross validation~\cite{Tib1,
Tib2}. Using this uncertainty, the simplest model is in principle obtained by
the 1-$\sigma$ rule, i.e., the maximal $\lambda$ compatible with the uncertainty
of $\hat\lambda_{\rm opt}$~\cite{Tib1, Tib2}. For the numerical simulations,  we
simply choose one value of $\hat\lambda_{\rm opt}=0.28$ for the penalty $P_2$, 
because uncertainties are dominated by the statistics of the lattice
measurements.  As mentioned above, the value $\hat\lambda_{\rm opt}=0.28$
corresponds to one  randomly chosen fit from the Monte-Carlo ensemble, but we
have made sure that this  value is representative.


\section{Partial twisting}\label{app:twisting}

In this section, we would like to examine in detail the unconventional twisting
prescription, which was introduced in Ref.~\cite{Sasaki}, in the context of
studying $J/\psi\phi$ scattering from $Y(4140)$ decays. We remind
the reader that, within this prescription, only quark fields are twisted,
whereas the antiquark fields are subject to the periodic boundary conditions.
One could ask whether such a prescription is rigorously justified.

We address this problem by using the same methods as in 
Ref.~\cite{Agadjanov-twisted}. In order to simplify things, we restrict
ourselves to the case of elastic $J/\psi\phi$ scattering and neglect
the coupling to the inelastic channels. In order to treat the partial twisting,
we introduce valence (${\sf v}$), sea (${\sf s}$) and ghost (${\sf g}$) quarks
for each quark flavor, subject to twisting. Only valence and ghost quarks are
twisted, whereas the sea  quarks are not. In total, 9 different
$J/\psi\phi$ states are possible
\begin{align}
1)&~(c_{\sf v}\bar c_{\sf v})\,(s_{\sf v}\bar s_{\sf v}) &
2)&~(c_{\sf v}\bar c_{\sf v})\,(s_{\sf s}\bar s_{\sf s})& 
3)&~(c_{\sf v}\bar c_{\sf v})\,(s_{\sf g}\bar s_{\sf g})&\nonumber\\
4)&~(c_{\sf s}\bar c_{\sf s})\,(s_{\sf v}\bar s_{\sf v}) &
5)&~(c_{\sf s}\bar c_{\sf s})\,(s_{\sf s}\bar s_{\sf s})& 
6)&~(c_{\sf s}\bar c_{\sf s})\,(s_{\sf g}\bar s_{\sf g})& \\
7)&~(c_{\sf g}\bar c_{\sf g})\,(s_{\sf v}\bar s_{\sf v}) & 
8)&~(c_{\sf g}\bar c_{\sf g})\,(s_{\sf s}\bar s_{\sf s})& 
9)&~(c_{\sf g}\bar c_{\sf g})\,(s_{\sf g}\bar s_{\sf g}) \ . \nonumber
\end{align}
The free Green's function is given by a a diagonal $9\times 9$ matrix. Taking
into account the sign convention for the mesons containing ghost quarks, this
matrix can be written in the following form
\eq
G=\mbox{diag}\,(G^\theta,G^+,-G^\theta,G^-,G^0,-G^-,-G^\theta,-G^+,G^\theta)\, .
\en
Here,
\eq
G^\theta({\bf p}_1,{\bf p}_2)&=&\frac{1}{2w_{J/\psi}({\bf p}_1+{\bf p}_\theta)
2w_\phi({\bf p}_2-{\bf p}_\theta)}\,
\frac{1}{w_{J/\psi}({\bf p}_1+{\bf p}_\theta)
+w_\phi({\bf p}_2-{\bf p}_\theta)-E}\, ,
\nonumber\\[2mm]
G^+({\bf p}_1,{\bf p}_2)&=&\frac{1}{2w_{J/\psi}({\bf p}_1+{\bf p}_\theta)
2w_\phi({\bf p}_2)}\,
\frac{1}{w_{J/\psi}({\bf p}_1+{\bf p}_\theta)+w_\phi({\bf p}_2)-E}\, ,
\nonumber\\[2mm]
G^-({\bf p}_1,{\bf p}_2)&=&\frac{1}{2w_{J/\psi}({\bf p}_1)
2w_\phi({\bf p}_2-{\bf p}_\theta)}\,
\frac{1}{w_{J/\psi}({\bf p}_1)+w_\phi({\bf p}_2-{\bf p}_\theta)-E}\, ,
\nonumber\\[2mm]
G^0({\bf p}_1,{\bf p}_2)&=&\frac{1}{2w_{J/\psi}({\bf p}_1)
2w_\phi({\bf p}_2)}\,
\frac{1}{w_{J/\psi}({\bf p}_1)+w_\phi({\bf p}_2)-E},
\en
where ${\bf p}_\theta=\boldsymbol{\theta}/L$ and
${\bf p}_i=2\pi/L\,{\bf n}_i\, , ~{\bf n}_i\in{\mathbb{Z}}^3\, , ~i=1,2$.

\begin{figure}[t]
\begin{center}
\includegraphics[width=14cm]{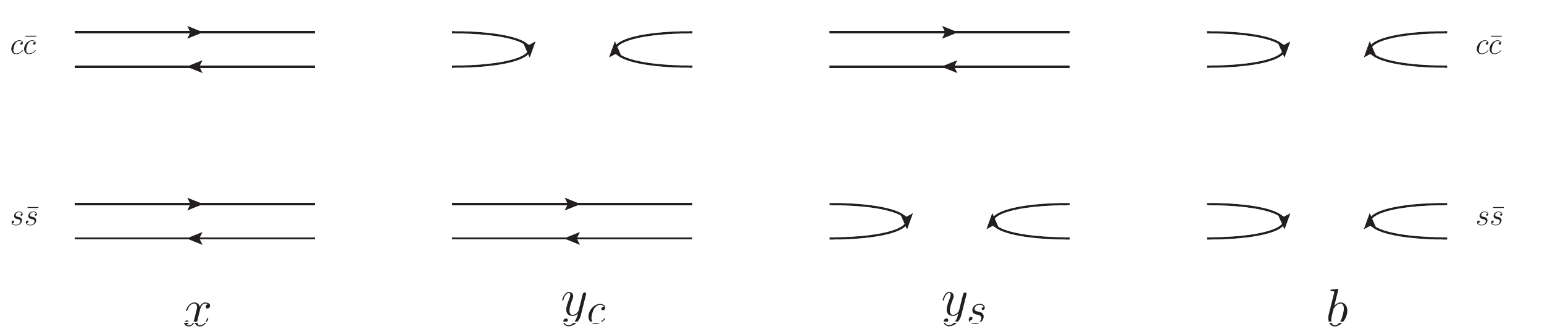}
\end{center}
\caption{The fully connected piece ($x$), the partially connected pieces 
($y_c$ and $y_s$) and the fully disconnected piece ($b$) of the $J/\psi\phi$ 
scattering amplitude.}
\label{fig:diagrams}
\end{figure}

The matrix elements that describe the transition of a state $i$ to state $j$,
$i,j=1,\ldots,9$ are given by
\eq\label{eq:Tmatrix}
T=\begin{pmatrix}
a & s & s & c & b & b & c & b & b \\
s & a & s & b & c & b & b & c & b \\
s & s & -a+2s & b & b & -c+2b & b & b & -c+2b \\
c & b & b & a & s & s & c & b & b \\
b & c & b & s & a & s & b & c & b \\
b & b & -c+2b & s & s & -a+2s & b & b & -c+2b \\
c & b & b & c & b & b & -a+2c & -s+2b & -s+2b \\
b & c & b & b & c & b & -s+2b & -a+2c & -s+2b \\
b & b & -c+2b & b & b & -c+2b & -s+2b & -s+2b & a-2c-2s+4b \\
\end{pmatrix}\, ,
\en
where 
\eq
a=x+y_c+y_s+b\, ,\quad\quad
s=y_s+b\, ,\quad\quad
c=y_c+b\, .
\en
The quantities $x,y_c,y_s,b$ denote the fully connected, partially connected and
fully disconnected contributions, see Fig.~\ref{fig:diagrams}. It is
straightforward to verify that the potential matrix $V$ in the infinite volume 
has exactly the same symmetries as the scattering matrix and is also given by
Eq.~(\ref{eq:Tmatrix}) with the replacement $a,b,c,s\to\tilde a,\tilde b,\tilde
c,\tilde s$.

The L\"uscher equation is given by
\eq\label{eq:Luescher-twisted}
\det(1-VG)&=&\ell_1^4\ell_2^2\ell_3^2\ell_4=0\, ,
\nonumber\\[2mm]
\ell_1&=&1-\langle G^\theta\rangle (\tilde a+\tilde b-\tilde c-\tilde s)\, ,
\nonumber\\[2mm]
\ell_2&=&1-\langle G^-\rangle(\tilde a-\tilde s)\, ,
\nonumber\\[2mm]
\ell_3&=&1-\langle G^+\rangle(\tilde a-\tilde c)\, ,
\nonumber\\[2mm]
\ell_4&=&1-\langle G^0\rangle \tilde a\, ,
\en
where
\eq
\langle G^\theta\rangle=\frac{1}{L^3}\sum_{\bf p}G^\theta({\bf p},-{\bf p})\, ,
\quad\quad
\langle G^0\rangle=\frac{1}{L^3}\sum_{\bf p}G^0({\bf p},-{\bf p})
\en
and $\langle G^\pm\rangle=0$ due to the conservation of the total momentum, 
if $\boldsymbol{\theta}$ is not equal to a multiple of $2\pi$.

As seen from Eq.~(\ref{eq:Luescher-twisted}), the finite-volume scattering
matrix at $\boldsymbol{\theta}\neq {\bf 0}$ contains two towers of poles,
determined by the equations $\ell_1=0$ and $\ell_4=0$, respectively,  where the
former depends on the parameter $\boldsymbol{\theta}$ and the latter  does not.
The explicit expression of the scattering matrix element in the valence sector
is given by
\eq\label{eq:vv-vv}
&&\left(V(1-GV)^{-1}\right)_{\sf vv,vv}
=\frac{\tilde a+\tilde b-\tilde s-\tilde c}{\ell_1}
+\frac{\tilde b^2}{\tilde a}\,\frac{1}{\ell_1^2\ell_4}
+\frac{2(\tilde b-\tilde c)(\tilde b-\tilde s)}{\tilde a+\tilde b
-\tilde c-\tilde s}\,\frac{1}{\ell_1^3}\nonumber\\[2mm]
&&+\,\frac{-\tilde b^3+(\tilde c+\tilde s)\tilde b^2-\tilde a^2\tilde b
+\tilde a^2(\tilde c+\tilde s)-\tilde a(\tilde c^2+\tilde s^2)
-4\tilde a(\tilde b-\tilde c)(\tilde b-\tilde s)}{\tilde a(\tilde a
+\tilde b-\tilde c-\tilde s)}\,\frac{1}{\ell_1^2}\, .
\nonumber\\
\en
It is also  clear that the $\boldsymbol{\theta}$-dependent singularities are 
determined by the fully connected part of the scattering amplitude, whereas the
$\boldsymbol{\theta}$-independent part contains the full amplitude.
Consequently, the approach of Ref.~\cite{Sasaki} can be safely used if and  only
if the contribution of the disconnected diagrams is much smaller than the
connected one (in fact, this was mentioned already in Ref.~\cite{Sasaki}). In
this case, i.e., when $\tilde b=\tilde c=\tilde s=0$,  the double and triple
poles in Eq.~(\ref{eq:vv-vv}) vanish and one arrives  at the expression that was
expected from the beginning 
\eq
\left(V(1-GV)^{-1}\right)_{\sf vv,vv}=\frac{\tilde a}
{1-\langle G^\theta \rangle \tilde a}\, .
\en
For the particular problem, considered here, one expects that the disconnected
contributions will be strongly suppressed, according to the OZI rule.
Consequently, the justification of the method, proposed in Ref.~\cite{Sasaki},
heavily rests on the effectiveness of the OZI suppression.


\end{document}